\documentclass[aps,prl,twocolumn,amsmath,amssymb]{revtex4-2}

\usepackage{graphicx}
\usepackage{color}
\usepackage[hidelinks]{hyperref}
\usepackage{orcidlink}

\hypersetup{colorlinks=true, linkcolor=black, citecolor=black, urlcolor=blue}

\begin{document}
\author{Karol Gietka\,\orcidlink{0000-0001-7700-3208}}
\email[]{karol.gietka@uibk.ac.at}
\author{Christoph Hotter\,\orcidlink{0009-0003-3854-0264}}
\author{Helmut Ritsch\,\orcidlink{0000-0001-7013-5208}}

\affiliation{Institut f\"ur Theoretische Physik, Universit\"at Innsbruck, A-6020 Innsbruck, Austria} 

\title{Unique Steady-State Squeezing in a Driven Quantum Rabi Model}

%%%%%%%%%%%%%%%%%%%%%%%%%%%%%%%%%%%%%%%%%%%%%%%%%%%%%%%%%%%%%%%%%%%%%%%%%%%%
%%%%%%%%%%%%%%%%%%%%%%%%%%%%%%%  ABSTRACT  %%%%%%%%%%%%%%%%%%%%%%%%%%%%%%%%%
%%%%%%%%%%%%%%%%%%%%%%%%%%%%%%%%%%%%%%%%%%%%%%%%%%%%%%%%%%%%%%%%%%%%%%%%%%%%

\begin{abstract}
Squeezing is essential to many quantum technologies and our understanding of quantum physics. Here we show a novel type of steady-state squeezing that can be generated in the closed and open quantum Rabi as well as Dicke model. To this end, we eliminate the spin dynamics which effectively leads to an abstract harmonic oscillator whose eigenstates are squeezed with respect to the non-interacting harmonic oscillator. By driving the system, we generate squeezing which has the unique property of time-independent uncertainties and squeezed dynamics. Such squeezing might find applications in continuous back-action evading measurements and should already be observable in optomechanical systems and Coulomb crystals.

\end{abstract}
\maketitle

%%%%%%%%%%%%%%%%%%%%%%%%%%%%%%%%%%%%%%%%%%%%%%%%%%%%%%%%%%%%%%%%%%%%%%%%%%%%
%%%%%%%%%%%%%%%%%%%%%%%%%%%%%%%  INTRODUCTION  %%%%%%%%%%%%%%%%%%%%%%%%%%%%%
%%%%%%%%%%%%%%%%%%%%%%%%%%%%%%%%%%%%%%%%%%%%%%%%%%%%%%%%%%%%%%%%%%%%%%%%%%%%

\emph{Introduction}. Squeezing~\cite{gardiner2004quantum,Zubairy_2005_squeezing} relies on redistributing quantum uncertainties between two non-commuting observables. The primary example is the squeezing of light~\cite{squeezing1983walls}, where the uncertainties are redistributed between the strength of electric and magnetic fields with respect to a coherent state where the uncertainties are equal. Squeezing is a precious quantum resource as it is rather robust to decoherence and dissipation. For this reason, it finds applications in many quantum technologies with the most prominent ones being high-precision measurements~\cite{laporta1987squeezedpolariztion,Kimble1987precision,polzik1992squeezing,geo6002011squeezing,grote2013firstsqueezing,SCHNABEL2017squeezing,ligo2019enhanced,virgo2019increasingsqueezeed,ligovirgo2021new,pan2023robustbeyondNOON, pedrozo2020entanglement, colombo2022timereversal} and entanglement-based quantum key distribution~\cite{schnabel2015CVkeydist,yongmin2019cvsqueezedqkd,Derkach_2020squeezingQKD,timothy2022squeezedlaser}. On the other hand, squeezing is a form of quantum correlation, important in the context of quantum phase transitions~\cite{plenio2015QRMphasetrans} and is used to study the fundamental aspects of quantum physics~\cite{koy2018squeezingbelltest}.

The quantum Rabi model~\cite{Braak_2016qrmcelebration80,Xie_2017rabimodelsolutionanddynamics} is a paradigmatic model in physics that describes a quantized harmonic oscillator coupled to a two-level system and its Hamiltonian reads ($\hbar=1$)
\begin{align}
    \hat H = \omega \hat a^\dagger \hat a + \frac{\Omega}{2}\hat \sigma_z + \frac{g}{2}(\hat a + \hat a^\dagger)\hat \sigma_x,
    \label{eq:H_rabi}
\end{align}
where $\hat a$ and $\hat a^\dagger$ are the annihilation and creation operators for a harmonic oscillator with frequency $\omega$. The Pauli matrices $\hat \sigma_z$ and $\hat \sigma_x$ describe the two-level system (here interchangeably referred to as spin) with frequency $\Omega$ and $g$ is the interaction strength between the two sub-systems. The interaction term can be rewritten with the help of spin raising and lowering operators, $\hat \sigma_x =(\hat \sigma_+ + \hat \sigma_-)$, into two terms 
\begin{align}
(\hat a + \hat a^\dagger)\hat \sigma_x = (\hat a \hat \sigma_- + \hat a^\dagger \hat \sigma_+) + (\hat a \hat \sigma_+ + \hat a^\dagger \hat \sigma_-).
\end{align}
The first one is typically referred to as the counter-rotating term and the second one is the rotating term. 
Neglecting the fast oscillating counter-rotating term leads to the Jaynes-Cummings model~\cite{knight1993JCmodel,JCbook2021} which is the backbone of modern quantum optics. This (rotating wave) approximation is valid for $g\ll \omega,\Omega$ and {$|\Omega-\omega|\ll|\Omega+\omega|$}, however, it is not able to capture every aspect of the rich and intriguing physics close to the critical point of the quantum Rabi model ($g\sim g_c \equiv \sqrt{\omega \Omega}$)~\cite{plenio2015QRMphasetrans} (for a detailed analysis of the critical behavior under the rotating wave approximation see Ref.~\cite{plenio2016JCcritical}).

In order to see why the vicinity of the critical point is interesting, we eliminate the dynamics of the spin using the Schrieffer-Wolff transformation~\cite{2011schriefferwolff}. Under the assumption of $1 -{g^2}/{g_c^2} \gg  \left(\omega/\Omega\right)^{2/3}$~\cite{gietka2022comqm}, this leads to
\begin{align}\label{eq:effRabi}
    \hat H_a = \omega \hat a^\dagger \hat a - \frac{g^2}{4 \Omega}\left(\hat a + \hat a^\dagger\right)^2
\end{align}
which is a squeezing Hamiltonian with eigenstates
\begin{align}
    |\psi_n \rangle = \exp\bigg\{\frac{1}{2}\left(\xi^*\hat a^2-\xi\hat a^{\dagger2}\right)\bigg\}|n\rangle,
\end{align} 
where $\xi = \frac{1}{4} \ln\{1-g^2/g_c^2\}$ is the squeezing parameter and $|n\rangle$ are the Fock states. Note that the above Hamiltonian can be diagonalized by introducing an operator $\hat c = \cosh(\xi)\hat a + \sinh(\xi) \hat a^\dagger$ which characterizes the eigenmode of a strongly interacting system with frequency $\omega\sqrt{1-g^2/g_c^2}$~\cite{[{See Supplemental Material at }][{ for details.}]sup1}. In other words, the spin can be thought of as a mediator of interactions between harmonic oscillator excitations in the limit of $\omega\ll\Omega $. 
\begin{figure*}[htb!]
    \centering
    \includegraphics[width=\textwidth]{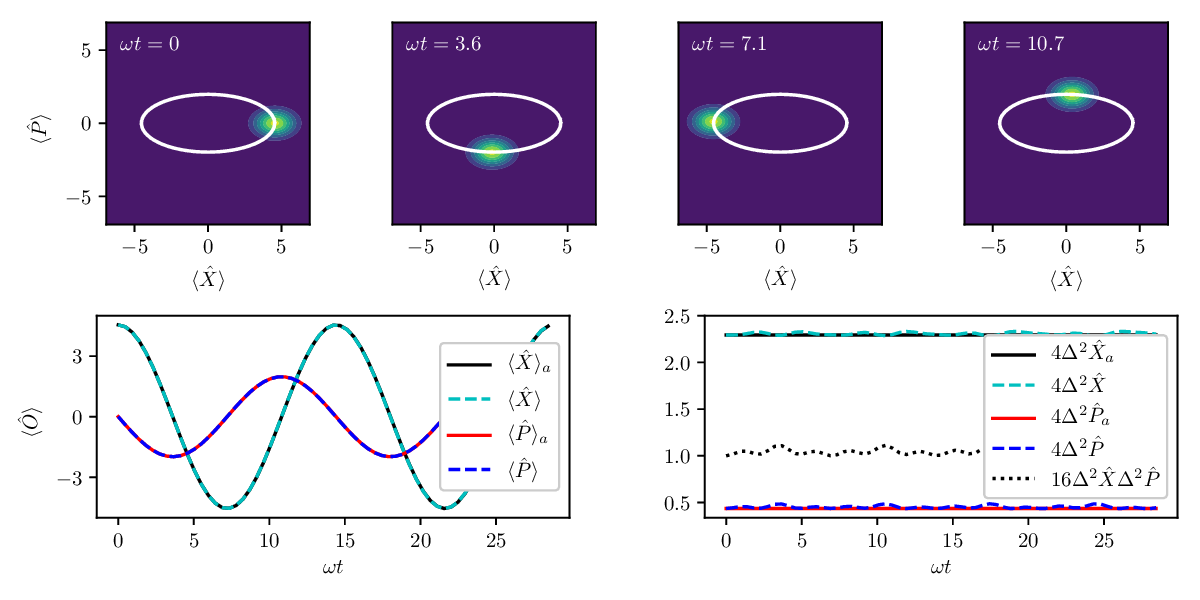}   
    \caption{Time evolution of the squeezed state for the kicked quantum Rabi model. The top panel shows the Husimi Q function at various times including the phase-space trajectory. The bottom left panel depicts the mean values $\langle\hat X \rangle$ and $\langle \hat P \rangle$ (subscript $a$ indicates the approximated abstract oscillator Hamiltonian) and the bottom right panel the squeezing of $\hat X$ and $\hat P$. The orbit (white line) is equally squeezed as the time-independent squeezed uncertainties. The dynamics described by the {\color{black} full quantum Rabi} (dashed lines) and the effective (solid lines) Hamiltonian agree very well, the visible wiggles for the squeezing appear because the simulation parameters are on the verge of the approximation breakdown. The parameters are $\Omega/\omega = 10^5$, $g/g_c = 0.9$, and $\alpha = 3$.} 
    \label{fig:fig1}
\end{figure*}

To interpret this squeezing we introduce the operators $\hat x = (\hat a + \hat a^\dagger)/\sqrt{2\omega}$ and $\hat p = \sqrt{\omega}(\hat a - \hat a^\dagger)/\sqrt{-2}$ {such that}
\begin{align}
    \hat H_a = \frac{\hat p^2}{2} + \frac{\omega^2}{2}\left(1-\frac{g^2}{g_c^2}\right)\hat x^2.
\end{align}
This Hamiltonian describes an abstract harmonic oscillator with a modified frequency $ \omega\sqrt{1-g^2/g_c^2}$ and a unit mass. Approaching then the critical point amounts to opening the abstract harmonic oscillator which leads to squeezing with respect to the non-interacting harmonic oscillator~\cite{gietka2022squeezing} $\omega \hat a^\dagger \hat a$ (the ground state of a given harmonic oscillator is a squeezed ground state of a harmonic oscillator with a different frequency). This means that if the measurement is performed in the basis of the non-interacting harmonic oscillator described by $\hat x$ and $\hat p$, increasing the coupling will lead to redistribution of uncertainties between $\hat x$ and $\hat p$. Therefore, the lower the effective frequency, the larger the spread $\Delta \hat x$. In particular, for $g=g_c$ the spread $\Delta \hat x$ is infinite and $\Delta \hat p$ becomes $0$ as the Hamiltonian describes an abstract free particle whose eigenstates are that of the abstract momentum. 

In this manuscript, we present how a driven (and dissipative) quantum Rabi model close to the critical point can generate (steady-state) squeezing of the harmonic oscillator excitations and its dynamics. Such squeezing has the unique property of time-independent uncertainties which might be crucial for several quantum technologies. We start with the closed quantum Rabi model, subsequently, we consider an open and driven system, and finally, we show how increasing the number of spins (Dicke model) leads to enhanced squeezing. We conclude by identifying potential applications of the presented squeezing mechanism and discussing possible platforms for the implementation of the protocol.

%%%%%%%%%%%%%%%%%%%%%%%%%%%%%%%%%%%%%%%%%%%%%%%%%%%%%%%%%%%%%%%%%%%%%%%%%%%%
%%%%%%%%%%%%%%%%%%%%%%%  KICKED QUANTUM RABI MODEL  %%%%%%%%%%%%%%%%%%%%%%%%
%%%%%%%%%%%%%%%%%%%%%%%%%%%%%%%%%%%%%%%%%%%%%%%%%%%%%%%%%%%%%%%%%%%%%%%%%%%%

\begin{figure*}[htb!]
    \centering
    \includegraphics[width=\textwidth]{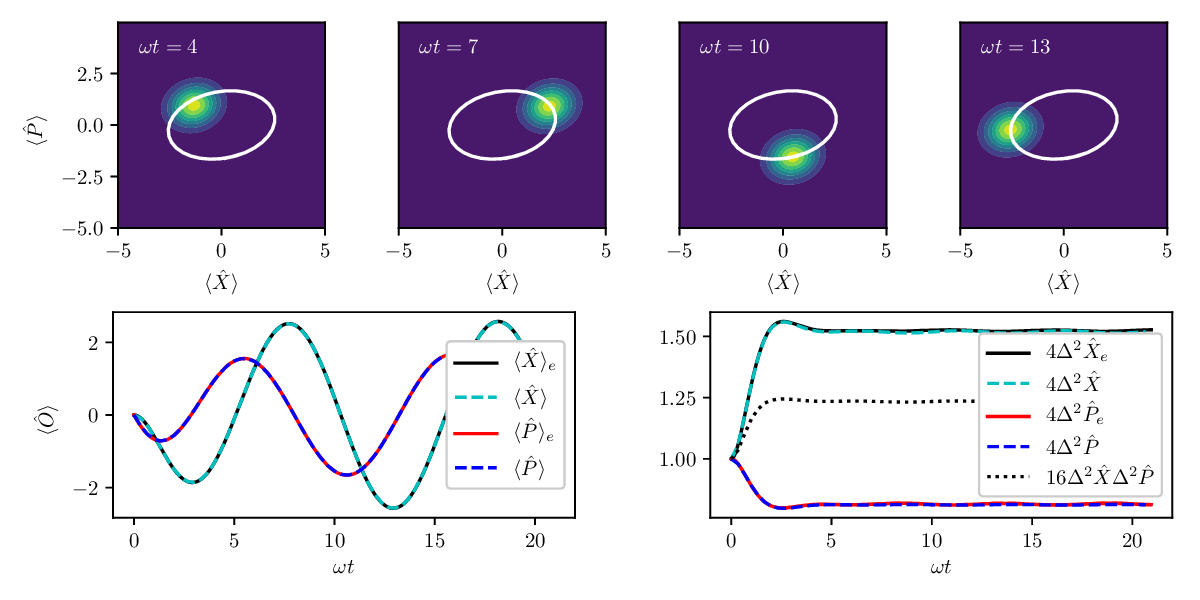}   
    \caption{Time evolution of the squeezed state for the driven-dissipative quantum Rabi model. The top panel shows the Husimi Q function at various times including the phase-space trajectory. The bottom left panel depicts the mean values $\langle\hat X \rangle$ and $\langle \hat P \rangle$ (subscript $e$ indicates the adiabatically eliminated abstract oscillator Hamiltonian) and the bottom right panel the squeezing of $\hat X$ and $\hat P$. The uncertainties reach a steady state (no need for the adiabatic time evolution) with only minor time dependence. The dynamics described by the {\color{black} full quantum Rabi} (dashed lines) and the effective (solid lines) Hamiltonian agree very well. The parameters are $\Omega/\omega = 2\cdot10^3$, $g/g_c = 0.8$, $\omega_d/\omega = \sqrt{1-g^2/g_c^2}$, $\eta/\omega =  1$ and $\kappa/\omega =  1$.}
    \label{fig:fig2}
\end{figure*}

\emph{Kicked Quantum Rabi model}. In order to excite the harmonic oscillator, we include a drive term to the quantum Rabi Hamiltonian \eqref{eq:H_rabi}
\begin{align}
    \hat H_d =   \eta \left(\hat a e^{i \omega_d t}  + \hat a^\dagger  e^{-i \omega_d t} \right),
\end{align}
where $\eta$ is the strength of the drive and $\omega_d$ is its frequency.  Although such a driving term is characteristic of laser-pumped cavities~\cite{Helmut2013rmpcavity} it can also describe driving of other harmonic oscillators. In an isolated system, the drive will excite the system indefinitely, therefore we assume a strong short pulse (kick) in this case. Such a drive acts as a displacement operator
$
    \hat D(\alpha) = \exp(\hat a \alpha + \hat a^\dagger \alpha^*)
$
which displaces an initial vacuum state by $\alpha$, creating a coherent state $|\alpha \rangle$. If the coupling strength $g$ is equal to $0$, the coherent state will rotate around the origin of the phase space with frequency $\omega$ at a fixed radius $|\alpha|$. Adiabatically increasing the coupling strength towards the critical point will then change the frequency of the abstract harmonic oscillator, leading to the change of the orbit from circular to elliptical and will redistribute the uncertainties between the quadrature operators $\hat X = (\hat a + \hat a^\dagger)/2$ and $\hat P = (\hat a - \hat a^\dagger)/2i$ (see Fig.~\ref{fig:fig1}). The final state can be easily found by constructing a coherent state out of squeezed Fock states of the harmonic mode $\hat a$. The time evolution of such a squeezed coherent state becomes then
\begin{align}
    |\psi (t) \rangle =e^{\frac{-|\alpha|^2}{2}} \sum_{n=0}^\infty e^{-i n \omega \sqrt{1-\frac{g^2}{g_c^2}} t}\frac{\alpha^n}{\sqrt{n!}}e^{\frac{1}{2}\left(\xi^* \hat a^2-\xi \hat a^{\dagger2} \right)}|n \rangle.
\end{align}

The equation of the new (squeezed) phase-space orbit can be found by calculating the average value of $\hat X$ and~$\hat P$
\begin{align}\label{eq:squeezedorb}
\begin{split}
        \langle \hat X \rangle =& \alpha\cos\left(\omega \sqrt{1-\frac{g^2}{g_c^2}} t\right)\exp(-\xi), \\ 
    \langle \hat P \rangle =& -\alpha\sin\left(\omega \sqrt{1-\frac{g^2}{g_c^2}} t\right)\exp(\xi).
\end{split}
\end{align}
The uncertainties in $\hat X$ and $\hat P$ can be calculated in the same way
\begin{align}\label{eq:squeezedunc}
\begin{split}
    \Delta^2\hat X &= \frac{1}{4}\exp(-2\xi), \\
    \Delta^2\hat P &= \frac{1}{4}\exp(2\xi),
    \end{split}
\end{align}
and turn out to be squeezed and time-independent. In contrast, creating squeezing in other ways, for example with a position measurement, squeezes the state but not the harmonic oscillator. Therefore, the squeezing experiences additional rotation which leads, for instance, to measurement back-action~\cite{lopez1995qndbackation,sillanpaa2016qbem,nunnenkamp2019squeezingbackation,mitchell2021backationevaging} and consequently to the standard quantum limit of measurement precision~\cite{caves1980rmpbackation}. In our approach, we adiabatically squeeze the abstract harmonic oscillator by changing its frequency which also squeezes the state. As a result, the state does not experience any rotations around its own axis at the \emph{expense} of exhibiting equally squeezed orbit and uncertainties [see Fig.~\ref{fig:fig1}, Eqs~\eqref{eq:squeezedorb} and \eqref{eq:squeezedunc}].

\begin{figure*}[htb!]
    \centering
    \includegraphics[width=\textwidth]{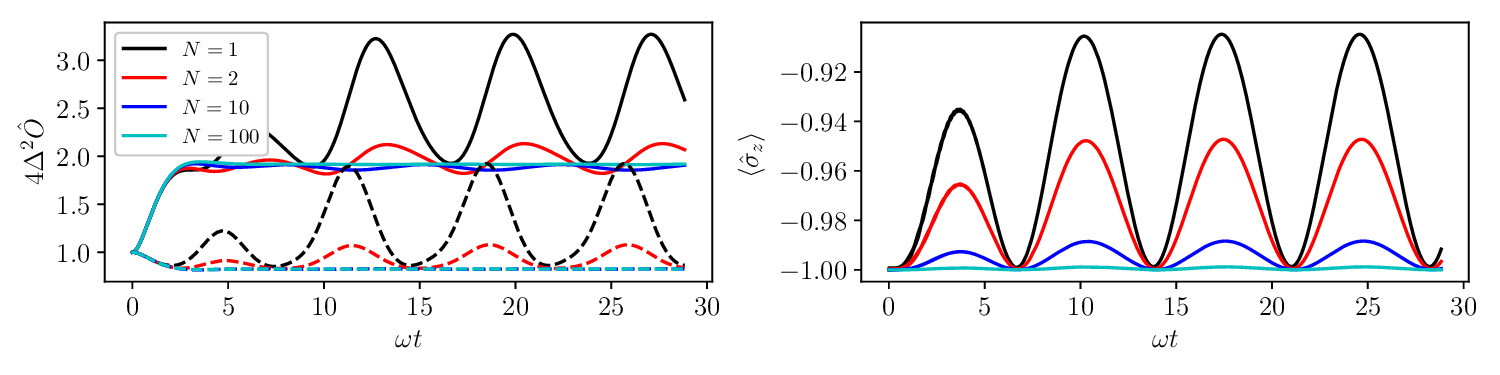}   
    \caption{Steady state squeezing for large number of spins. The left panel shows the squeezing of $\hat X$ (solid lines) and $\hat P$ (dashed lines) and the right side shows the excitation of one spin. For large number of spins $N$ the excitation per spin $\langle \hat{\sigma}_z \rangle$ is decreased and the uncertainties reach a steady state (cyan). The parameters are $\Omega/\omega = 2\cdot10^3$, $g/g_c = 0.9$, $\omega_d/\omega = \sqrt{1-g^2/g_c^2}$, $\eta/\omega =  4$ and $\kappa/\omega =  1$.}
    \label{fig:fig3}
\end{figure*}

In practice, obtaining large and detectable squeezing of a macroscopic coherent state ($| \alpha | \gg 1$) might be difficult with a single two-level system, as it would require an extremely large ratio $\Omega/\omega$ to prevent the spin from getting excited (the effective Hamiltonian will no longer be valid). One way to artificially increase the atomic frequency in the quantum Rabi model is by enlarging the number of spins (Dicke model) which can be naively understood as changing the frequency from $\Omega$ to $N \Omega$~\cite{gietka2021invertedoscDicke}, with $N$ being the number of two-level systems (see Fig. \ref{fig:fig3}). 
In the limiting case $N\rightarrow \infty$, the large spin can be replaced by another harmonic oscillator by means of the Holstein-Primakoff transformation~\cite{1940HPtransf}.

%%%%%%%%%%%%%%%%%%%%%%%%%%%%%%%%%%%%%%%%%%%%%%%%%%%%%%%%%%%%%%%%%%%%%%%%%%%%
%%%%%%%%%%%%%%%%%  DRIVEN-DISSIPATIVE QUANTUM RABI MODEL  %%%%%%%%%%%%%%%%%%
%%%%%%%%%%%%%%%%%%%%%%%%%%%%%%%%%%%%%%%%%%%%%%%%%%%%%%%%%%%%%%%%%%%%%%%%%%%%

\emph{Driven-Dissipative Quantum Rabi model}. In an open quantum system, the dissipation will eventually bring the state of the system to the ground state. In order to prevent this, we continuously drive the system.  Since the effective Hamiltonian is quadratic the system is described by a Gaussian state, and we expect that the dynamics of the {\color{black} full quantum Rabi} Hamiltonian is also well approximated by Gaussian physics. Therefore, we use a second-order mean-field description \cite{Plankensteiner2022quantumcumulantsjl}, which leads to a closed set of equations~\cite{[{See the \textsc{QuantumCumulants.jl} code at }][{ for details.}]supp}. The effect of %spin and 
harmonic oscillator dissipation with rate $\kappa$ is taken into account by means of the Lindblad master equation
\begin{align}
\begin{split}
\mathcal{L}[\hat \rho] = &  -i\left[\hat H + \hat H_d, \hat \rho \right]   +\kappa\left(\hat c \hat \rho \hat c^\dagger -\frac{1}{2}\big\{\hat c^\dagger\hat c, \hat \rho \big\}\right).
\end{split}
\end{align}
Note that once the coupling is strong, the jump operators have to be redefined~\cite{dissipation2011blais,rabl2018dissipativeQRM,Cattaneo_2019,RevModPhys.91.025005}. The role of the jump operators for a dissipative process is to bring the (un-driven) system to a unique ground state. For strongly interacting systems, the ground state has much more energy than the ground state of a non-interacting system. In other words, the ground state of strongly interacting systems is a highly excited state of a non-interacting system. Therefore, using naively the jump operators for the non-interacting systems would lead to unphysical behavior as, for instance, extracting energy from the ground state~\cite{[{See Supplemental Material at }][{ for details.}]sup1}. A similar argument holds for the driving term. Once the coupling is strong, we are no longer driving the bare mode described by $\hat a$ but a new (dressed) mode described by $\hat c$. 
% {\color{blue} I would not write that much here, only reference to the supplement with one sentence or so. I wrote something a little bit below.}
% {\color{black} In the Lindblad master equation, we do not consider dissipation of the spin degree of freedom. Since the protocol requires the spin to be in its ground state, it might seem that the dissipation of the spin degree of freedom is always beneficial. This, however, is only true for weak dissipation. For strongly damped systems, the energy uncertainty of the spin degree of freedom will prevent from generating the steady-state squeezing as indicated by numerical calculations~\cite{[{See Supplemental Material at }][{ for details.}]sup1}}. 

After adiabatic elimination of the spin dynamics (equivalent to the Schrieffer-Wolff transformation for the closed system), we can find the effective Hamiltonian
\begin{align} \label{eq:effac}
    \hat H_{e} = \omega \hat a^\dagger \hat a - \frac{g^2}{4 \Omega}\left(\hat a + \hat a^\dagger\right)^2 + \eta (\hat c e^{i \omega_d t} + \hat c^\dagger e^{-i \omega_d t}),
\end{align}
which describes the abstract harmonic oscillator with an additional modified drive term. Fig. \ref{fig:fig2} depicts the time evolution of the squeezing for the driven dissipative case. At the top and lower left panel we see that the introduction of drive and dissipation does not change the results qualitatively. In the driven-dissipative case the amplitude of the phase space oscillations is related to $2\eta/\kappa$. The lower right panel shows the almost time-independent squeezing after some settling time. The time evolution described by the adiabatically eliminated Hamiltonian (solid lines) agrees very well with that of the {\color{black} full quantum Rabi} Hamiltonian (dashed lines) {\color{black} including the spin degree of freedom}. For non-negligible spin excitation this approximation breaks down, furthermore, also the uncertainties become strongly time-dependent. This can be suppressed by increasing the number of spins $N$. In Fig~\ref{fig:fig3}, we show the time dependence of the squeezing (left) and spin excitation (right) for different $N$. For a sufficiently large number of spins the uncertainties become time-independent. 
{\color{black} Including weak dissipation of the spin can also suppress the excitation and lead to time-independent uncertainties. However, strong damping can prevent the system from generating squeezing~\cite{[{See Supplemental Material at }][{ for details.}]sup1}.}

We would like to point out that the direction of squeezing depends on the type of coupling $g(\hat a e^{i \theta} + \hat a^\dagger e^{-i \theta})\hat \sigma_x$ and it is related to $\theta$ (this also holds for the closed system). For instance, choosing $\theta$ to be $\pi/2$ would result in a harmonic oscillator where the $\hat X$ quadrature is squeezed and not anti-squeezed as in this work. In principle, by adjusting $\theta$ one could obtain squeezing in an arbitrary quadrature direction. This in turn suggests that it should be possible to observe motion of a harmonic oscillator where at the point of maximal displacement the oscillator has the maximal momentum, for example, by tilting the axis by $\pi/4$. Such behavior is unfathomable for a classical harmonic oscillator. 

%%%%%%%%%%%%%%%%%%%%%%%%%%%%%%%%%%%%%%%%%%%%%%%%%%%%%%%%%%%%%%%%%%%%%%%%%%%%
%%%%%%%%%%%%%%%%%%%%%%%%%%%%%% MEASUREMENT %%%%%%%%%%%%%%%%%%%%%%%%%%%%%%%%%
%%%%%%%%%%%%%%%%%%%%%%%%%%%%%%%%%%%%%%%%%%%%%%%%%%%%%%%%%%%%%%%%%%%%%%%%%%%%

\emph{Squeezing detection}. So far the description was general, and we did not specify the harmonic oscillator and the underlying physical system. In this section, we discuss whether it is possible to observe steady-state squeezing and how to do it. In the case of a mechanical oscillator whose internal degree of freedom is coupled to the center-of-mass motion (phonon mode), the squeezing could be simply observed by measuring the position or momentum of the center of mass. In this case, the squeezing manifests itself in decreased or increased uncertainty of position and momentum. Such squeezing should already be realizable in optomechanical systems~\cite{uros2020,uros2022} and ions interacting with a common phonon mode~\cite{2015ionCoulombCrystal,2019CrystalDickesym} by weakly driving the ground state close to the critical point. The crucial element that allows to observe squeezing in these systems is the well-defined observables $\hat x$ and $\hat p$ for the uncoupled mechanical oscillator.

For the electromagnetic resonators, the measurement of squeezing is different. In this case, the squeezing manifests itself in a changed resonance frequency of the resonator. This means that the definition of $\hat x$ and $\hat p$ depends on the frequency of the resonator. Therefore it is impossible to perform a measurement in the basis of the non-interacting harmonic oscillator~\cite{[{See Supplemental Material at }][{ for details.}]sup1}. The cavity in which atoms are strongly coupled to a single mode of radiation cannot generate squeezed light simply by driving it close to the critical point of the phase transition. 
From another perspective, light-matter interactions change the index of refraction and hence the resonance frequency which can be understood as squeezing of the electromagnetic field by changing its frequency.

%%%%%%%%%%%%%%%%%%%%%%%%%%%%%%%%%%%%%%%%%%%%%%%%%%%%%%%%%%%%%%%%%%%%%%%%%%%%
%%%%%%%%%%%%%%%%%%%%%%%%  DISCUSSION & CONCLUSIONS %%%%%%%%%%%%%%%%%%%%%%%%%
%%%%%%%%%%%%%%%%%%%%%%%%%%%%%%%%%%%%%%%%%%%%%%%%%%%%%%%%%%%%%%%%%%%%%%%%%%%%

\emph{Conclusions}. We have presented unique steady-state squeezing in the closed and open quantum Rabi model. In both cases, we obtain steady-state squeezing with time-independent variances and a squeezed trajectory in the phase-space picture defined by the non-interacting harmonic oscillator. Such squeezing can find applications in many quantum technologies, in particular, in quantum back-action-free continuous measurements~\cite{caves1980rmpbackation,schwab2010nanoQBFmeasure,polzik2017baemeasure} and driven-dissipative~\cite{porras2023ddmetrologycollectivespin,ilias2023criticalityenhanced} critical~\cite{2020criticalwitkowska,Gietka2021adiabaticcritical,2021liuexperiment,2021criticalwitkowska,2022PRXQuantumcontinus,2022_Garbe_heisenbegkible,Gietka2022understanding,2022entropycritical,2022dingnature,Aybar2022criticalquantum,2022garbeexponetialsqueezing,gietka2023squeezingHeisenberg} metrology. In order to understand squeezing, we introduced an effective Hamiltonian describing an abstract harmonic oscillator which we obtain by eliminating the dynamics of the two-level system. A promising extension of this proposal is the possibility of driving the spin directly and eliminating subsequently its dynamics in the dispersive regime to see steady-state squeezing. 

Since a quantum harmonic oscillator coupled to a two-level (or multiple-level) system can be used to describe many systems, the proposed method could be tested in a variety of physical platforms including mechanical resonators~\cite{2013clerk,aspelmeyer2014cavityoptom,uros2020,uros2022,dare2023linear}, spin-orbit coupled quantum gases~\cite{busch2016socreview,gietka2022comqm}, ions coupled to phonons~\cite{duan2021qrbphonon}, Coulomb crystals~\cite{2015ionCoulombCrystal,2019CrystalDickesym}, and even electrons trapped on a surface of liquid helium~\cite{konstantinov2019qrmhelium}. We predict, that the most promising system for the implementation of the described steady-state squeezing would be linear optomechanical systems in the far red-detuned and ultrastrong coupling regime~\cite{kustura2022mechanicalsqueezing}. Also, systems composed of $N$ trapped ions interacting with a single phononic mode (realizing the Dicke model) might be a perfect platform to create steady-state squeezing as it should be relatively easy to enter the regime where the effective Hamiltonian is valid. 

In principle, it should also be possible to observe steady-state squeezing on the other side of the critical point $g > g_c$ where the system can be described by a slightly modified effective Hamiltonian \eqref{eq:effRabi} (see Ref.~\cite{plenio2015QRMphasetrans} for details). From a practical point of view, however, the generation of such squeezing would require an extremely large detuning $\Omega \gg \omega$ and would generate a macroscopic number of excitations. Also, beyond the critical point, the ground state is double degenerate and we expect that this degeneracy further hinders the possibility of observing the steady-state squeezing with constant variances.

%%%%%%%%%%%%%%%%%%%%%%%%%%%%%%%%%%%%%%%%%%%%%%%%%%%%%%%%%%%%%%%%%%%%%%%%%%%%
%%%%%%%%%%%%%%%%%%%%%%%%%%% ACKNOWLEDGEMENTS %%%%%%%%%%%%%%%%%%%%%%%%%%%%%%%
%%%%%%%%%%%%%%%%%%%%%%%%%%%%%%%%%%%%%%%%%%%%%%%%%%%%%%%%%%%%%%%%%%%%%%%%%%%%

\emph{Acknowledgements}.
We would like to acknowledge Gerhard Kirchmair and Nico Baßler for discussions. {Simulations were performed using the open-source frameworks \textsc{QuantumOptics.jl}~\cite{kramer2018quantumoptics} and \textsc{QuantumCumulants.jl}~\cite{Plankensteiner2022quantumcumulantsjl}.} This work was supported by the Lise-Meitner Fellowship M3304-N of the Austrian Science Fund (FWF).

% \bibliography{bibliography.bib}

\begin{thebibliography}{72}%
\makeatletter
\providecommand \@ifxundefined [1]{%
 \@ifx{#1\undefined}
}%
\providecommand \@ifnum [1]{%
 \ifnum #1\expandafter \@firstoftwo
 \else \expandafter \@secondoftwo
 \fi
}%
\providecommand \@ifx [1]{%
 \ifx #1\expandafter \@firstoftwo
 \else \expandafter \@secondoftwo
 \fi
}%
\providecommand \natexlab [1]{#1}%
\providecommand \enquote  [1]{``#1''}%
\providecommand \bibnamefont  [1]{#1}%
\providecommand \bibfnamefont [1]{#1}%
\providecommand \citenamefont [1]{#1}%
\providecommand \href@noop [0]{\@secondoftwo}%
\providecommand \href [0]{\begingroup \@sanitize@url \@href}%
\providecommand \@href[1]{\@@startlink{#1}\@@href}%
\providecommand \@@href[1]{\endgroup#1\@@endlink}%
\providecommand \@sanitize@url [0]{\catcode `\\12\catcode `\$12\catcode
  `\&12\catcode `\#12\catcode `\^12\catcode `\_12\catcode `\%12\relax}%
\providecommand \@@startlink[1]{}%
\providecommand \@@endlink[0]{}%
\providecommand \url  [0]{\begingroup\@sanitize@url \@url }%
\providecommand \@url [1]{\endgroup\@href {#1}{\urlprefix }}%
\providecommand \urlprefix  [0]{URL }%
\providecommand \Eprint [0]{\href }%
\providecommand \doibase [0]{https://doi.org/}%
\providecommand \selectlanguage [0]{\@gobble}%
\providecommand \bibinfo  [0]{\@secondoftwo}%
\providecommand \bibfield  [0]{\@secondoftwo}%
\providecommand \translation [1]{[#1]}%
\providecommand \BibitemOpen [0]{}%
\providecommand \bibitemStop [0]{}%
\providecommand \bibitemNoStop [0]{.\EOS\space}%
\providecommand \EOS [0]{\spacefactor3000\relax}%
\providecommand \BibitemShut  [1]{\csname bibitem#1\endcsname}%
\let\auto@bib@innerbib\@empty
%</preamble>
\bibitem [{\citenamefont {Gardiner}\ and\ \citenamefont
  {Zoller}(2004)}]{gardiner2004quantum}%
  \BibitemOpen
  \bibfield  {author} {\bibinfo {author} {\bibfnamefont {C.}~\bibnamefont
  {Gardiner}}\ and\ \bibinfo {author} {\bibfnamefont {P.}~\bibnamefont
  {Zoller}},\ }\href@noop {} {\emph {\bibinfo {title} {Quantum noise: a
  handbook of Markovian and non-Markovian quantum stochastic methods with
  applications to quantum optics}}}\ (\bibinfo  {publisher} {Springer Science
  \& Business Media},\ \bibinfo {year} {2004})\BibitemShut {NoStop}%
\bibitem [{\citenamefont {Zubairy}(2005)}]{Zubairy_2005_squeezing}%
  \BibitemOpen
  \bibfield  {author} {\bibinfo {author} {\bibfnamefont {S.}~\bibnamefont
  {Zubairy}},\ }\bibfield  {title} {\bibinfo {title} {Quantum squeezing},\
  }\href {https://doi.org/10.1088/1464-4266/7/5/B01} {\bibfield  {journal}
  {\bibinfo  {journal} {Journal of Optics B: Quantum and Semiclassical Optics}\
  }\textbf {\bibinfo {volume} {7}},\ \bibinfo {pages} {156} (\bibinfo {year}
  {2005})}\BibitemShut {NoStop}%
\bibitem [{\citenamefont {Walls}(1983)}]{squeezing1983walls}%
  \BibitemOpen
  \bibfield  {author} {\bibinfo {author} {\bibfnamefont {D.~F.}\ \bibnamefont
  {Walls}},\ }\bibfield  {title} {\bibinfo {title} {Squeezed states of light},\
  }\href {https://doi.org/10.1038/306141a0} {\bibfield  {journal} {\bibinfo
  {journal} {Nature}\ }\textbf {\bibinfo {volume} {306}},\ \bibinfo {pages}
  {141} (\bibinfo {year} {1983})}\BibitemShut {NoStop}%
\bibitem [{\citenamefont {Grangier}\ \emph {et~al.}(1987)\citenamefont
  {Grangier}, \citenamefont {Slusher}, \citenamefont {Yurke},\ and\
  \citenamefont {LaPorta}}]{laporta1987squeezedpolariztion}%
  \BibitemOpen
  \bibfield  {author} {\bibinfo {author} {\bibfnamefont {P.}~\bibnamefont
  {Grangier}}, \bibinfo {author} {\bibfnamefont {R.~E.}\ \bibnamefont
  {Slusher}}, \bibinfo {author} {\bibfnamefont {B.}~\bibnamefont {Yurke}},\
  and\ \bibinfo {author} {\bibfnamefont {A.}~\bibnamefont {LaPorta}},\
  }\bibfield  {title} {\bibinfo {title} {Squeezed-light--enhanced polarization
  interferometer},\ }\href {https://doi.org/10.1103/PhysRevLett.59.2153}
  {\bibfield  {journal} {\bibinfo  {journal} {Phys. Rev. Lett.}\ }\textbf
  {\bibinfo {volume} {59}},\ \bibinfo {pages} {2153} (\bibinfo {year}
  {1987})}\BibitemShut {NoStop}%
\bibitem [{\citenamefont {Xiao}\ \emph {et~al.}(1987)\citenamefont {Xiao},
  \citenamefont {Wu},\ and\ \citenamefont {Kimble}}]{Kimble1987precision}%
  \BibitemOpen
  \bibfield  {author} {\bibinfo {author} {\bibfnamefont {M.}~\bibnamefont
  {Xiao}}, \bibinfo {author} {\bibfnamefont {L.-A.}\ \bibnamefont {Wu}},\ and\
  \bibinfo {author} {\bibfnamefont {H.~J.}\ \bibnamefont {Kimble}},\ }\bibfield
   {title} {\bibinfo {title} {Precision measurement beyond the shot-noise
  limit},\ }\href {https://doi.org/10.1103/PhysRevLett.59.278} {\bibfield
  {journal} {\bibinfo  {journal} {Phys. Rev. Lett.}\ }\textbf {\bibinfo
  {volume} {59}},\ \bibinfo {pages} {278} (\bibinfo {year} {1987})}\BibitemShut
  {NoStop}%
\bibitem [{\citenamefont {Polzik}\ \emph {et~al.}(1992)\citenamefont {Polzik},
  \citenamefont {Carri},\ and\ \citenamefont {Kimble}}]{polzik1992squeezing}%
  \BibitemOpen
  \bibfield  {author} {\bibinfo {author} {\bibfnamefont {E.~S.}\ \bibnamefont
  {Polzik}}, \bibinfo {author} {\bibfnamefont {J.}~\bibnamefont {Carri}},\ and\
  \bibinfo {author} {\bibfnamefont {H.~J.}\ \bibnamefont {Kimble}},\ }\bibfield
   {title} {\bibinfo {title} {Spectroscopy with squeezed light},\ }\href
  {https://doi.org/10.1103/PhysRevLett.68.3020} {\bibfield  {journal} {\bibinfo
   {journal} {Phys. Rev. Lett.}\ }\textbf {\bibinfo {volume} {68}},\ \bibinfo
  {pages} {3020} (\bibinfo {year} {1992})}\BibitemShut {NoStop}%
\bibitem [{\citenamefont {{LIGO Scientific
  Collaboration}}(2011)}]{geo6002011squeezing}%
  \BibitemOpen
  \bibfield  {author} {\bibinfo {author} {\bibnamefont {{LIGO Scientific
  Collaboration}}},\ }\bibfield  {title} {\bibinfo {title} {A gravitational
  wave observatory operating beyond the quantum shot-noise limit},\ }\href
  {https://doi.org/10.1038/nphys2083} {\bibfield  {journal} {\bibinfo
  {journal} {Nature Physics}\ }\textbf {\bibinfo {volume} {7}},\ \bibinfo
  {pages} {962} (\bibinfo {year} {2011})}\BibitemShut {NoStop}%
\bibitem [{\citenamefont {Grote}\ \emph {et~al.}(2013)\citenamefont {Grote},
  \citenamefont {Danzmann}, \citenamefont {Dooley}, \citenamefont {Schnabel},
  \citenamefont {Slutsky},\ and\ \citenamefont
  {Vahlbruch}}]{grote2013firstsqueezing}%
  \BibitemOpen
  \bibfield  {author} {\bibinfo {author} {\bibfnamefont {H.}~\bibnamefont
  {Grote}}, \bibinfo {author} {\bibfnamefont {K.}~\bibnamefont {Danzmann}},
  \bibinfo {author} {\bibfnamefont {K.~L.}\ \bibnamefont {Dooley}}, \bibinfo
  {author} {\bibfnamefont {R.}~\bibnamefont {Schnabel}}, \bibinfo {author}
  {\bibfnamefont {J.}~\bibnamefont {Slutsky}},\ and\ \bibinfo {author}
  {\bibfnamefont {H.}~\bibnamefont {Vahlbruch}},\ }\bibfield  {title} {\bibinfo
  {title} {First long-term application of squeezed states of light in a
  gravitational-wave observatory},\ }\href
  {https://doi.org/10.1103/PhysRevLett.110.181101} {\bibfield  {journal}
  {\bibinfo  {journal} {Phys. Rev. Lett.}\ }\textbf {\bibinfo {volume} {110}},\
  \bibinfo {pages} {181101} (\bibinfo {year} {2013})}\BibitemShut {NoStop}%
\bibitem [{\citenamefont {Schnabel}(2017)}]{SCHNABEL2017squeezing}%
  \BibitemOpen
  \bibfield  {author} {\bibinfo {author} {\bibfnamefont {R.}~\bibnamefont
  {Schnabel}},\ }\bibfield  {title} {\bibinfo {title} {Squeezed states of light
  and their applications in laser interferometers},\ }\href
  {https://doi.org/https://doi.org/10.1016/j.physrep.2017.04.001} {\bibfield
  {journal} {\bibinfo  {journal} {Physics Reports}\ }\textbf {\bibinfo {volume}
  {684}},\ \bibinfo {pages} {1} (\bibinfo {year} {2017})},\ \bibinfo {note}
  {squeezed states of light and their applications in laser
  interferometers}\BibitemShut {NoStop}%
\bibitem [{\citenamefont {{LIGO Scientific
  Collaboration}}(2019)}]{ligo2019enhanced}%
  \BibitemOpen
  \bibfield  {author} {\bibinfo {author} {\bibnamefont {{LIGO Scientific
  Collaboration}}},\ }\bibfield  {title} {\bibinfo {title} {Quantum-enhanced
  advanced ligo detectors in the era of gravitational-wave astronomy},\ }\href
  {https://doi.org/10.1103/PhysRevLett.123.231107} {\bibfield  {journal}
  {\bibinfo  {journal} {Phys. Rev. Lett.}\ }\textbf {\bibinfo {volume} {123}},\
  \bibinfo {pages} {231107} (\bibinfo {year} {2019})}\BibitemShut {NoStop}%
\bibitem [{\citenamefont {{Virgo
  Collaboration}}(2019)}]{virgo2019increasingsqueezeed}%
  \BibitemOpen
  \bibfield  {author} {\bibinfo {author} {\bibnamefont {{Virgo
  Collaboration}}},\ }\bibfield  {title} {\bibinfo {title} {Increasing the
  astrophysical reach of the advanced virgo detector via the application of
  squeezed vacuum states of light},\ }\href
  {https://doi.org/10.1103/PhysRevLett.123.231108} {\bibfield  {journal}
  {\bibinfo  {journal} {Phys. Rev. Lett.}\ }\textbf {\bibinfo {volume} {123}},\
  \bibinfo {pages} {231108} (\bibinfo {year} {2019})}\BibitemShut {NoStop}%
\bibitem [{\citenamefont {{LIGO Scientific Collaboration and Virgo
  Collaboration}}(2021)}]{ligovirgo2021new}%
  \BibitemOpen
  \bibfield  {author} {\bibinfo {author} {\bibnamefont {{LIGO Scientific
  Collaboration and Virgo Collaboration}}},\ }\bibfield  {title} {\bibinfo
  {title} {Gwtc-2: Compact binary coalescences observed by ligo and virgo
  during the first half of the third observing run},\ }\href
  {https://doi.org/10.1103/PhysRevX.11.021053} {\bibfield  {journal} {\bibinfo
  {journal} {Phys. Rev. X}\ }\textbf {\bibinfo {volume} {11}},\ \bibinfo
  {pages} {021053} (\bibinfo {year} {2021})}\BibitemShut {NoStop}%
\bibitem [{\citenamefont {Qin}\ \emph {et~al.}(2023)\citenamefont {Qin},
  \citenamefont {Deng}, \citenamefont {Zhong}, \citenamefont {Peng},
  \citenamefont {Su}, \citenamefont {Luo}, \citenamefont {Xu}, \citenamefont
  {Wu}, \citenamefont {Gong}, \citenamefont {Liu}, \citenamefont {Wang},
  \citenamefont {Chen}, \citenamefont {Li}, \citenamefont {Liu}, \citenamefont
  {Lu},\ and\ \citenamefont {Pan}}]{pan2023robustbeyondNOON}%
  \BibitemOpen
  \bibfield  {author} {\bibinfo {author} {\bibfnamefont {J.}~\bibnamefont
  {Qin}}, \bibinfo {author} {\bibfnamefont {Y.-H.}\ \bibnamefont {Deng}},
  \bibinfo {author} {\bibfnamefont {H.-S.}\ \bibnamefont {Zhong}}, \bibinfo
  {author} {\bibfnamefont {L.-C.}\ \bibnamefont {Peng}}, \bibinfo {author}
  {\bibfnamefont {H.}~\bibnamefont {Su}}, \bibinfo {author} {\bibfnamefont
  {Y.-H.}\ \bibnamefont {Luo}}, \bibinfo {author} {\bibfnamefont {J.-M.}\
  \bibnamefont {Xu}}, \bibinfo {author} {\bibfnamefont {D.}~\bibnamefont {Wu}},
  \bibinfo {author} {\bibfnamefont {S.-Q.}\ \bibnamefont {Gong}}, \bibinfo
  {author} {\bibfnamefont {H.-L.}\ \bibnamefont {Liu}}, \bibinfo {author}
  {\bibfnamefont {H.}~\bibnamefont {Wang}}, \bibinfo {author} {\bibfnamefont
  {M.-C.}\ \bibnamefont {Chen}}, \bibinfo {author} {\bibfnamefont
  {L.}~\bibnamefont {Li}}, \bibinfo {author} {\bibfnamefont {N.-L.}\
  \bibnamefont {Liu}}, \bibinfo {author} {\bibfnamefont {C.-Y.}\ \bibnamefont
  {Lu}},\ and\ \bibinfo {author} {\bibfnamefont {J.-W.}\ \bibnamefont {Pan}},\
  }\bibfield  {title} {\bibinfo {title} {Unconditional and robust quantum
  metrological advantage beyond n00n states},\ }\href
  {https://doi.org/10.1103/PhysRevLett.130.070801} {\bibfield  {journal}
  {\bibinfo  {journal} {Phys. Rev. Lett.}\ }\textbf {\bibinfo {volume} {130}},\
  \bibinfo {pages} {070801} (\bibinfo {year} {2023})}\BibitemShut {NoStop}%
\bibitem [{\citenamefont {Pedrozo-Peñafiel}\ \emph {et~al.}(2020)\citenamefont
  {Pedrozo-Peñafiel}, \citenamefont {Colombo}, \citenamefont {Shu},
  \citenamefont {Adiyatullin}, \citenamefont {Li}, \citenamefont {Mendez},
  \citenamefont {Braverman}, \citenamefont {Kawasaki}, \citenamefont
  {Akamatsu}, \citenamefont {Xiao},\ and\ \citenamefont
  {Vuletić}}]{pedrozo2020entanglement}%
  \BibitemOpen
  \bibfield  {author} {\bibinfo {author} {\bibfnamefont {E.}~\bibnamefont
  {Pedrozo-Peñafiel}}, \bibinfo {author} {\bibfnamefont {S.}~\bibnamefont
  {Colombo}}, \bibinfo {author} {\bibfnamefont {C.}~\bibnamefont {Shu}},
  \bibinfo {author} {\bibfnamefont {A.~F.}\ \bibnamefont {Adiyatullin}},
  \bibinfo {author} {\bibfnamefont {Z.}~\bibnamefont {Li}}, \bibinfo {author}
  {\bibfnamefont {E.}~\bibnamefont {Mendez}}, \bibinfo {author} {\bibfnamefont
  {B.}~\bibnamefont {Braverman}}, \bibinfo {author} {\bibfnamefont
  {A.}~\bibnamefont {Kawasaki}}, \bibinfo {author} {\bibfnamefont
  {D.}~\bibnamefont {Akamatsu}}, \bibinfo {author} {\bibfnamefont
  {Y.}~\bibnamefont {Xiao}},\ and\ \bibinfo {author} {\bibfnamefont
  {V.}~\bibnamefont {Vuletić}},\ }\bibfield  {title} {\bibinfo {title}
  {Entanglement on an optical atomic-clock transition},\ }\bibfield  {journal}
  {\bibinfo  {journal} {Nature}\ }\textbf {\bibinfo {volume} {588}},\ \href
  {https://doi.org/10.1038/s41586-020-3006-1} {10.1038/s41586-020-3006-1}
  (\bibinfo {year} {2020})\BibitemShut {NoStop}%
\bibitem [{\citenamefont {Colombo}\ \emph {et~al.}(2022)\citenamefont
  {Colombo}, \citenamefont {Pedrozo-Peñafiel}, \citenamefont {Adiyatullin},
  \citenamefont {Li}, \citenamefont {Mendez}, \citenamefont {Shu},\ and\
  \citenamefont {Vuletić}}]{colombo2022timereversal}%
  \BibitemOpen
  \bibfield  {author} {\bibinfo {author} {\bibfnamefont {S.}~\bibnamefont
  {Colombo}}, \bibinfo {author} {\bibfnamefont {E.}~\bibnamefont
  {Pedrozo-Peñafiel}}, \bibinfo {author} {\bibfnamefont {A.~F.}\ \bibnamefont
  {Adiyatullin}}, \bibinfo {author} {\bibfnamefont {Z.}~\bibnamefont {Li}},
  \bibinfo {author} {\bibfnamefont {E.}~\bibnamefont {Mendez}}, \bibinfo
  {author} {\bibfnamefont {C.}~\bibnamefont {Shu}},\ and\ \bibinfo {author}
  {\bibfnamefont {V.}~\bibnamefont {Vuletić}},\ }\bibfield  {title} {\bibinfo
  {title} {Time-reversal-based quantum metrology with many-body entangled
  states},\ }\href {https://doi.org/10.1038/s41567-022-01653-5} {\bibfield
  {journal} {\bibinfo  {journal} {Nature Physics}\ }\textbf {\bibinfo {volume}
  {18}},\ \bibinfo {pages} {925–930} (\bibinfo {year} {2022})}\BibitemShut
  {NoStop}%
\bibitem [{\citenamefont {Gehring}\ \emph {et~al.}(2015)\citenamefont
  {Gehring}, \citenamefont {H{\"a}ndchen}, \citenamefont {Duhme}, \citenamefont
  {Furrer}, \citenamefont {Franz}, \citenamefont {Pacher}, \citenamefont
  {Werner},\ and\ \citenamefont {Schnabel}}]{schnabel2015CVkeydist}%
  \BibitemOpen
  \bibfield  {author} {\bibinfo {author} {\bibfnamefont {T.}~\bibnamefont
  {Gehring}}, \bibinfo {author} {\bibfnamefont {V.}~\bibnamefont
  {H{\"a}ndchen}}, \bibinfo {author} {\bibfnamefont {J.}~\bibnamefont {Duhme}},
  \bibinfo {author} {\bibfnamefont {F.}~\bibnamefont {Furrer}}, \bibinfo
  {author} {\bibfnamefont {T.}~\bibnamefont {Franz}}, \bibinfo {author}
  {\bibfnamefont {C.}~\bibnamefont {Pacher}}, \bibinfo {author} {\bibfnamefont
  {R.~F.}\ \bibnamefont {Werner}},\ and\ \bibinfo {author} {\bibfnamefont
  {R.}~\bibnamefont {Schnabel}},\ }\bibfield  {title} {\bibinfo {title}
  {Implementation of continuous-variable quantum key distribution with
  composable and one-sided-device-independent security against coherent
  attacks},\ }\href {https://doi.org/10.1038/ncomms9795} {\bibfield  {journal}
  {\bibinfo  {journal} {Nature Communications}\ }\textbf {\bibinfo {volume}
  {6}},\ \bibinfo {pages} {8795} (\bibinfo {year} {2015})}\BibitemShut
  {NoStop}%
\bibitem [{\citenamefont {Wang}\ \emph {et~al.}(2019)\citenamefont {Wang},
  \citenamefont {Wang},\ and\ \citenamefont {Li}}]{yongmin2019cvsqueezedqkd}%
  \BibitemOpen
  \bibfield  {author} {\bibinfo {author} {\bibfnamefont {P.}~\bibnamefont
  {Wang}}, \bibinfo {author} {\bibfnamefont {X.}~\bibnamefont {Wang}},\ and\
  \bibinfo {author} {\bibfnamefont {Y.}~\bibnamefont {Li}},\ }\bibfield
  {title} {\bibinfo {title} {Continuous-variable measurement-device-independent
  quantum key distribution using modulated squeezed states and optical
  amplifiers},\ }\href {https://doi.org/10.1103/PhysRevA.99.042309} {\bibfield
  {journal} {\bibinfo  {journal} {Phys. Rev. A}\ }\textbf {\bibinfo {volume}
  {99}},\ \bibinfo {pages} {042309} (\bibinfo {year} {2019})}\BibitemShut
  {NoStop}%
\bibitem [{\citenamefont {Derkach}\ \emph {et~al.}(2020)\citenamefont
  {Derkach}, \citenamefont {Usenko},\ and\ \citenamefont
  {Filip}}]{Derkach_2020squeezingQKD}%
  \BibitemOpen
  \bibfield  {author} {\bibinfo {author} {\bibfnamefont {I.}~\bibnamefont
  {Derkach}}, \bibinfo {author} {\bibfnamefont {V.~C.}\ \bibnamefont
  {Usenko}},\ and\ \bibinfo {author} {\bibfnamefont {R.}~\bibnamefont
  {Filip}},\ }\bibfield  {title} {\bibinfo {title} {Squeezing-enhanced quantum
  key distribution over atmospheric channels},\ }\href
  {https://doi.org/10.1088/1367-2630/ab7f8f} {\bibfield  {journal} {\bibinfo
  {journal} {New Journal of Physics}\ }\textbf {\bibinfo {volume} {22}},\
  \bibinfo {pages} {053006} (\bibinfo {year} {2020})}\BibitemShut {NoStop}%
\bibitem [{\citenamefont {Hosseinidehaj}\ \emph {et~al.}(2022)\citenamefont
  {Hosseinidehaj}, \citenamefont {Winnel},\ and\ \citenamefont
  {Ralph}}]{timothy2022squeezedlaser}%
  \BibitemOpen
  \bibfield  {author} {\bibinfo {author} {\bibfnamefont {N.}~\bibnamefont
  {Hosseinidehaj}}, \bibinfo {author} {\bibfnamefont {M.~S.}\ \bibnamefont
  {Winnel}},\ and\ \bibinfo {author} {\bibfnamefont {T.~C.}\ \bibnamefont
  {Ralph}},\ }\bibfield  {title} {\bibinfo {title} {Simple and loss-tolerant
  free-space quantum key distribution using a squeezed laser},\ }\href
  {https://doi.org/10.1103/PhysRevA.105.032602} {\bibfield  {journal} {\bibinfo
   {journal} {Phys. Rev. A}\ }\textbf {\bibinfo {volume} {105}},\ \bibinfo
  {pages} {032602} (\bibinfo {year} {2022})}\BibitemShut {NoStop}%
\bibitem [{\citenamefont {Hwang}\ \emph {et~al.}(2015)\citenamefont {Hwang},
  \citenamefont {Puebla},\ and\ \citenamefont
  {Plenio}}]{plenio2015QRMphasetrans}%
  \BibitemOpen
  \bibfield  {author} {\bibinfo {author} {\bibfnamefont {M.-J.}\ \bibnamefont
  {Hwang}}, \bibinfo {author} {\bibfnamefont {R.}~\bibnamefont {Puebla}},\ and\
  \bibinfo {author} {\bibfnamefont {M.~B.}\ \bibnamefont {Plenio}},\ }\bibfield
   {title} {\bibinfo {title} {Quantum phase transition and universal dynamics
  in the rabi model},\ }\href {https://doi.org/10.1103/PhysRevLett.115.180404}
  {\bibfield  {journal} {\bibinfo  {journal} {Phys. Rev. Lett.}\ }\textbf
  {\bibinfo {volume} {115}},\ \bibinfo {pages} {180404} (\bibinfo {year}
  {2015})}\BibitemShut {NoStop}%
\bibitem [{\citenamefont {Thearle}\ \emph {et~al.}(2018)\citenamefont
  {Thearle}, \citenamefont {Janousek}, \citenamefont {Armstrong}, \citenamefont
  {Hosseini}, \citenamefont {Sch\"unemann~(Mraz)}, \citenamefont {Assad},
  \citenamefont {Symul}, \citenamefont {James}, \citenamefont {Huntington},
  \citenamefont {Ralph},\ and\ \citenamefont {Lam}}]{koy2018squeezingbelltest}%
  \BibitemOpen
  \bibfield  {author} {\bibinfo {author} {\bibfnamefont {O.}~\bibnamefont
  {Thearle}}, \bibinfo {author} {\bibfnamefont {J.}~\bibnamefont {Janousek}},
  \bibinfo {author} {\bibfnamefont {S.}~\bibnamefont {Armstrong}}, \bibinfo
  {author} {\bibfnamefont {S.}~\bibnamefont {Hosseini}}, \bibinfo {author}
  {\bibfnamefont {M.}~\bibnamefont {Sch\"unemann~(Mraz)}}, \bibinfo {author}
  {\bibfnamefont {S.}~\bibnamefont {Assad}}, \bibinfo {author} {\bibfnamefont
  {T.}~\bibnamefont {Symul}}, \bibinfo {author} {\bibfnamefont {M.~R.}\
  \bibnamefont {James}}, \bibinfo {author} {\bibfnamefont {E.}~\bibnamefont
  {Huntington}}, \bibinfo {author} {\bibfnamefont {T.~C.}\ \bibnamefont
  {Ralph}},\ and\ \bibinfo {author} {\bibfnamefont {P.~K.}\ \bibnamefont
  {Lam}},\ }\bibfield  {title} {\bibinfo {title} {Violation of bell's
  inequality using continuous variable measurements},\ }\href
  {https://doi.org/10.1103/PhysRevLett.120.040406} {\bibfield  {journal}
  {\bibinfo  {journal} {Phys. Rev. Lett.}\ }\textbf {\bibinfo {volume} {120}},\
  \bibinfo {pages} {040406} (\bibinfo {year} {2018})}\BibitemShut {NoStop}%
\bibitem [{\citenamefont {Braak}\ \emph {et~al.}(2016)\citenamefont {Braak},
  \citenamefont {Chen}, \citenamefont {Batchelor},\ and\ \citenamefont
  {Solano}}]{Braak_2016qrmcelebration80}%
  \BibitemOpen
  \bibfield  {author} {\bibinfo {author} {\bibfnamefont {D.}~\bibnamefont
  {Braak}}, \bibinfo {author} {\bibfnamefont {Q.-H.}\ \bibnamefont {Chen}},
  \bibinfo {author} {\bibfnamefont {M.~T.}\ \bibnamefont {Batchelor}},\ and\
  \bibinfo {author} {\bibfnamefont {E.}~\bibnamefont {Solano}},\ }\bibfield
  {title} {\bibinfo {title} {Semi-classical and quantum rabi models: in
  celebration of 80 years},\ }\href
  {https://doi.org/10.1088/1751-8113/49/30/300301} {\bibfield  {journal}
  {\bibinfo  {journal} {Journal of Physics A: Mathematical and Theoretical}\
  }\textbf {\bibinfo {volume} {49}},\ \bibinfo {pages} {300301} (\bibinfo
  {year} {2016})}\BibitemShut {NoStop}%
\bibitem [{\citenamefont {Xie}\ \emph {et~al.}(2017)\citenamefont {Xie},
  \citenamefont {Zhong}, \citenamefont {Batchelor},\ and\ \citenamefont
  {Lee}}]{Xie_2017rabimodelsolutionanddynamics}%
  \BibitemOpen
  \bibfield  {author} {\bibinfo {author} {\bibfnamefont {Q.}~\bibnamefont
  {Xie}}, \bibinfo {author} {\bibfnamefont {H.}~\bibnamefont {Zhong}}, \bibinfo
  {author} {\bibfnamefont {M.~T.}\ \bibnamefont {Batchelor}},\ and\ \bibinfo
  {author} {\bibfnamefont {C.}~\bibnamefont {Lee}},\ }\bibfield  {title}
  {\bibinfo {title} {The quantum rabi model: solution and dynamics},\ }\href
  {https://doi.org/10.1088/1751-8121/aa5a65} {\bibfield  {journal} {\bibinfo
  {journal} {Journal of Physics A: Mathematical and Theoretical}\ }\textbf
  {\bibinfo {volume} {50}},\ \bibinfo {pages} {113001} (\bibinfo {year}
  {2017})}\BibitemShut {NoStop}%
\bibitem [{\citenamefont {Shore}\ and\ \citenamefont
  {Knight}(1993)}]{knight1993JCmodel}%
  \BibitemOpen
  \bibfield  {author} {\bibinfo {author} {\bibfnamefont {B.~W.}\ \bibnamefont
  {Shore}}\ and\ \bibinfo {author} {\bibfnamefont {P.~L.}\ \bibnamefont
  {Knight}},\ }\bibfield  {title} {\bibinfo {title} {The jaynes-cummings
  model},\ }\href {https://doi.org/10.1080/09500349314551321} {\bibfield
  {journal} {\bibinfo  {journal} {Journal of Modern Optics}\ }\textbf {\bibinfo
  {volume} {40}},\ \bibinfo {pages} {1195} (\bibinfo {year}
  {1993})}\BibitemShut {NoStop}%
\bibitem [{\citenamefont {Larson}\ and\ \citenamefont
  {Mavrogordatos}(2021)}]{JCbook2021}%
  \BibitemOpen
  \bibfield  {author} {\bibinfo {author} {\bibfnamefont {J.}~\bibnamefont
  {Larson}}\ and\ \bibinfo {author} {\bibfnamefont {T.}~\bibnamefont
  {Mavrogordatos}},\ }\href {https://doi.org/10.1088/978-0-7503-3447-1} {\emph
  {\bibinfo {title} {The Jaynes–Cummings Model and Its Descendants}}},\
  2053-2563\ (\bibinfo  {publisher} {IOP Publishing},\ \bibinfo {year}
  {2021})\BibitemShut {NoStop}%
\bibitem [{\citenamefont {Hwang}\ and\ \citenamefont
  {Plenio}(2016)}]{plenio2016JCcritical}%
  \BibitemOpen
  \bibfield  {author} {\bibinfo {author} {\bibfnamefont {M.-J.}\ \bibnamefont
  {Hwang}}\ and\ \bibinfo {author} {\bibfnamefont {M.~B.}\ \bibnamefont
  {Plenio}},\ }\bibfield  {title} {\bibinfo {title} {Quantum phase transition
  in the finite jaynes-cummings lattice systems},\ }\href
  {https://doi.org/10.1103/PhysRevLett.117.123602} {\bibfield  {journal}
  {\bibinfo  {journal} {Phys. Rev. Lett.}\ }\textbf {\bibinfo {volume} {117}},\
  \bibinfo {pages} {123602} (\bibinfo {year} {2016})}\BibitemShut {NoStop}%
\bibitem [{\citenamefont {Bravyi}\ \emph {et~al.}(2011)\citenamefont {Bravyi},
  \citenamefont {DiVincenzo},\ and\ \citenamefont
  {Loss}}]{2011schriefferwolff}%
  \BibitemOpen
  \bibfield  {author} {\bibinfo {author} {\bibfnamefont {S.}~\bibnamefont
  {Bravyi}}, \bibinfo {author} {\bibfnamefont {D.~P.}\ \bibnamefont
  {DiVincenzo}},\ and\ \bibinfo {author} {\bibfnamefont {D.}~\bibnamefont
  {Loss}},\ }\bibfield  {title} {\bibinfo {title} {Schrieffer–wolff
  transformation for quantum many-body systems},\ }\href
  {https://doi.org/https://doi.org/10.1016/j.aop.2011.06.004} {\bibfield
  {journal} {\bibinfo  {journal} {Annals of Physics}\ }\textbf {\bibinfo
  {volume} {326}},\ \bibinfo {pages} {2793} (\bibinfo {year}
  {2011})}\BibitemShut {NoStop}%
\bibitem [{\citenamefont {Gietka}(2022{\natexlab{a}})}]{gietka2022comqm}%
  \BibitemOpen
  \bibfield  {author} {\bibinfo {author} {\bibfnamefont {K.}~\bibnamefont
  {Gietka}},\ }\bibfield  {title} {\bibinfo {title} {Harnessing center-of-mass
  excitations in quantum metrology},\ }\href
  {https://doi.org/10.1103/PhysRevResearch.4.043074} {\bibfield  {journal}
  {\bibinfo  {journal} {Phys. Rev. Res.}\ }\textbf {\bibinfo {volume} {4}},\
  \bibinfo {pages} {043074} (\bibinfo {year} {2022}{\natexlab{a}})}\BibitemShut
  {NoStop}%
\bibitem [{sup({\natexlab{a}})}]{sup1}%
  \BibitemOpen
  \href@noop {} {}\bibinfo {howpublished}
  {\url{https://journals.aps.org/prl/supplemental/10.1103/PhysRevLett.131.223604/Supplemental_Material.pdf}} ({\natexlab{a}})\BibitemShut
  {NoStop}%
\bibitem [{\citenamefont {Gietka}(2022{\natexlab{b}})}]{gietka2022squeezing}%
  \BibitemOpen
  \bibfield  {author} {\bibinfo {author} {\bibfnamefont {K.}~\bibnamefont
  {Gietka}},\ }\bibfield  {title} {\bibinfo {title} {Squeezing by critical
  speeding up: Applications in quantum metrology},\ }\href
  {https://doi.org/10.1103/PhysRevA.105.042620} {\bibfield  {journal} {\bibinfo
   {journal} {Phys. Rev. A}\ }\textbf {\bibinfo {volume} {105}},\ \bibinfo
  {pages} {042620} (\bibinfo {year} {2022}{\natexlab{b}})}\BibitemShut
  {NoStop}%
\bibitem [{\citenamefont {Ritsch}\ \emph {et~al.}(2013)\citenamefont {Ritsch},
  \citenamefont {Domokos}, \citenamefont {Brennecke},\ and\ \citenamefont
  {Esslinger}}]{Helmut2013rmpcavity}%
  \BibitemOpen
  \bibfield  {author} {\bibinfo {author} {\bibfnamefont {H.}~\bibnamefont
  {Ritsch}}, \bibinfo {author} {\bibfnamefont {P.}~\bibnamefont {Domokos}},
  \bibinfo {author} {\bibfnamefont {F.}~\bibnamefont {Brennecke}},\ and\
  \bibinfo {author} {\bibfnamefont {T.}~\bibnamefont {Esslinger}},\ }\bibfield
  {title} {\bibinfo {title} {Cold atoms in cavity-generated dynamical optical
  potentials},\ }\href {https://doi.org/10.1103/RevModPhys.85.553} {\bibfield
  {journal} {\bibinfo  {journal} {Rev. Mod. Phys.}\ }\textbf {\bibinfo {volume}
  {85}},\ \bibinfo {pages} {553} (\bibinfo {year} {2013})}\BibitemShut
  {NoStop}%
\bibitem [{\citenamefont {Bencheikh}\ \emph {et~al.}(1995)\citenamefont
  {Bencheikh}, \citenamefont {Levenson}, \citenamefont {Grangier},\ and\
  \citenamefont {Lopez}}]{lopez1995qndbackation}%
  \BibitemOpen
  \bibfield  {author} {\bibinfo {author} {\bibfnamefont {K.}~\bibnamefont
  {Bencheikh}}, \bibinfo {author} {\bibfnamefont {J.~A.}\ \bibnamefont
  {Levenson}}, \bibinfo {author} {\bibfnamefont {P.}~\bibnamefont {Grangier}},\
  and\ \bibinfo {author} {\bibfnamefont {O.}~\bibnamefont {Lopez}},\ }\bibfield
   {title} {\bibinfo {title} {Quantum nondemolition demonstration via repeated
  backaction evading measurements},\ }\href
  {https://doi.org/10.1103/PhysRevLett.75.3422} {\bibfield  {journal} {\bibinfo
   {journal} {Phys. Rev. Lett.}\ }\textbf {\bibinfo {volume} {75}},\ \bibinfo
  {pages} {3422} (\bibinfo {year} {1995})}\BibitemShut {NoStop}%
\bibitem [{\citenamefont {Ockeloen-Korppi}\ \emph {et~al.}(2016)\citenamefont
  {Ockeloen-Korppi}, \citenamefont {Damsk\"agg}, \citenamefont {Pirkkalainen},
  \citenamefont {Clerk}, \citenamefont {Woolley},\ and\ \citenamefont
  {Sillanp\"a\"a}}]{sillanpaa2016qbem}%
  \BibitemOpen
  \bibfield  {author} {\bibinfo {author} {\bibfnamefont {C.~F.}\ \bibnamefont
  {Ockeloen-Korppi}}, \bibinfo {author} {\bibfnamefont {E.}~\bibnamefont
  {Damsk\"agg}}, \bibinfo {author} {\bibfnamefont {J.-M.}\ \bibnamefont
  {Pirkkalainen}}, \bibinfo {author} {\bibfnamefont {A.~A.}\ \bibnamefont
  {Clerk}}, \bibinfo {author} {\bibfnamefont {M.~J.}\ \bibnamefont {Woolley}},\
  and\ \bibinfo {author} {\bibfnamefont {M.~A.}\ \bibnamefont
  {Sillanp\"a\"a}},\ }\bibfield  {title} {\bibinfo {title} {Quantum backaction
  evading measurement of collective mechanical modes},\ }\href
  {https://doi.org/10.1103/PhysRevLett.117.140401} {\bibfield  {journal}
  {\bibinfo  {journal} {Phys. Rev. Lett.}\ }\textbf {\bibinfo {volume} {117}},\
  \bibinfo {pages} {140401} (\bibinfo {year} {2016})}\BibitemShut {NoStop}%
\bibitem [{\citenamefont {Brunelli}\ \emph {et~al.}(2019)\citenamefont
  {Brunelli}, \citenamefont {Malz},\ and\ \citenamefont
  {Nunnenkamp}}]{nunnenkamp2019squeezingbackation}%
  \BibitemOpen
  \bibfield  {author} {\bibinfo {author} {\bibfnamefont {M.}~\bibnamefont
  {Brunelli}}, \bibinfo {author} {\bibfnamefont {D.}~\bibnamefont {Malz}},\
  and\ \bibinfo {author} {\bibfnamefont {A.}~\bibnamefont {Nunnenkamp}},\
  }\bibfield  {title} {\bibinfo {title} {Conditional dynamics of optomechanical
  two-tone backaction-evading measurements},\ }\href
  {https://doi.org/10.1103/PhysRevLett.123.093602} {\bibfield  {journal}
  {\bibinfo  {journal} {Phys. Rev. Lett.}\ }\textbf {\bibinfo {volume} {123}},\
  \bibinfo {pages} {093602} (\bibinfo {year} {2019})}\BibitemShut {NoStop}%
\bibitem [{\citenamefont {Troullinou}\ \emph {et~al.}(2021)\citenamefont
  {Troullinou}, \citenamefont {Jim\'enez-Mart\'{\i}nez}, \citenamefont {Kong},
  \citenamefont {Lucivero},\ and\ \citenamefont
  {Mitchell}}]{mitchell2021backationevaging}%
  \BibitemOpen
  \bibfield  {author} {\bibinfo {author} {\bibfnamefont {C.}~\bibnamefont
  {Troullinou}}, \bibinfo {author} {\bibfnamefont {R.}~\bibnamefont
  {Jim\'enez-Mart\'{\i}nez}}, \bibinfo {author} {\bibfnamefont
  {J.}~\bibnamefont {Kong}}, \bibinfo {author} {\bibfnamefont {V.~G.}\
  \bibnamefont {Lucivero}},\ and\ \bibinfo {author} {\bibfnamefont {M.~W.}\
  \bibnamefont {Mitchell}},\ }\bibfield  {title} {\bibinfo {title}
  {Squeezed-light enhancement and backaction evasion in a high sensitivity
  optically pumped magnetometer},\ }\href
  {https://doi.org/10.1103/PhysRevLett.127.193601} {\bibfield  {journal}
  {\bibinfo  {journal} {Phys. Rev. Lett.}\ }\textbf {\bibinfo {volume} {127}},\
  \bibinfo {pages} {193601} (\bibinfo {year} {2021})}\BibitemShut {NoStop}%
\bibitem [{\citenamefont {Caves}\ \emph {et~al.}(1980)\citenamefont {Caves},
  \citenamefont {Thorne}, \citenamefont {Drever}, \citenamefont {Sandberg},\
  and\ \citenamefont {Zimmermann}}]{caves1980rmpbackation}%
  \BibitemOpen
  \bibfield  {author} {\bibinfo {author} {\bibfnamefont {C.~M.}\ \bibnamefont
  {Caves}}, \bibinfo {author} {\bibfnamefont {K.~S.}\ \bibnamefont {Thorne}},
  \bibinfo {author} {\bibfnamefont {R.~W.~P.}\ \bibnamefont {Drever}}, \bibinfo
  {author} {\bibfnamefont {V.~D.}\ \bibnamefont {Sandberg}},\ and\ \bibinfo
  {author} {\bibfnamefont {M.}~\bibnamefont {Zimmermann}},\ }\bibfield  {title}
  {\bibinfo {title} {On the measurement of a weak classical force coupled to a
  quantum-mechanical oscillator. i. issues of principle},\ }\href
  {https://doi.org/10.1103/RevModPhys.52.341} {\bibfield  {journal} {\bibinfo
  {journal} {Rev. Mod. Phys.}\ }\textbf {\bibinfo {volume} {52}},\ \bibinfo
  {pages} {341} (\bibinfo {year} {1980})}\BibitemShut {NoStop}%
\bibitem [{\citenamefont {Gietka}\ and\ \citenamefont
  {Busch}(2021)}]{gietka2021invertedoscDicke}%
  \BibitemOpen
  \bibfield  {author} {\bibinfo {author} {\bibfnamefont {K.}~\bibnamefont
  {Gietka}}\ and\ \bibinfo {author} {\bibfnamefont {T.}~\bibnamefont {Busch}},\
  }\bibfield  {title} {\bibinfo {title} {Inverted harmonic oscillator dynamics
  of the nonequilibrium phase transition in the dicke model},\ }\href
  {https://doi.org/10.1103/PhysRevE.104.034132} {\bibfield  {journal} {\bibinfo
   {journal} {Phys. Rev. E}\ }\textbf {\bibinfo {volume} {104}},\ \bibinfo
  {pages} {034132} (\bibinfo {year} {2021})}\BibitemShut {NoStop}%
\bibitem [{\citenamefont {Holstein}\ and\ \citenamefont
  {Primakoff}(1940)}]{1940HPtransf}%
  \BibitemOpen
  \bibfield  {author} {\bibinfo {author} {\bibfnamefont {T.}~\bibnamefont
  {Holstein}}\ and\ \bibinfo {author} {\bibfnamefont {H.}~\bibnamefont
  {Primakoff}},\ }\bibfield  {title} {\bibinfo {title} {Field dependence of the
  intrinsic domain magnetization of a ferromagnet},\ }\href
  {https://doi.org/10.1103/PhysRev.58.1098} {\bibfield  {journal} {\bibinfo
  {journal} {Phys. Rev.}\ }\textbf {\bibinfo {volume} {58}},\ \bibinfo {pages}
  {1098} (\bibinfo {year} {1940})}\BibitemShut {NoStop}%
\bibitem [{\citenamefont {Plankensteiner}\ \emph {et~al.}(2022)\citenamefont
  {Plankensteiner}, \citenamefont {Hotter},\ and\ \citenamefont
  {Ritsch}}]{Plankensteiner2022quantumcumulantsjl}%
  \BibitemOpen
  \bibfield  {author} {\bibinfo {author} {\bibfnamefont {D.}~\bibnamefont
  {Plankensteiner}}, \bibinfo {author} {\bibfnamefont {C.}~\bibnamefont
  {Hotter}},\ and\ \bibinfo {author} {\bibfnamefont {H.}~\bibnamefont
  {Ritsch}},\ }\bibfield  {title} {\bibinfo {title} {Quantum{C}umulants.jl: {A}
  {J}ulia framework for generalized mean-field equations in open quantum
  systems},\ }\href {https://dx.doi.org/10.22331/q-2022-01-04-617} {\bibfield
  {journal} {\bibinfo  {journal} {{Quantum}}\ }\textbf {\bibinfo {volume} {6}}
  (\bibinfo {year} {2022})}\BibitemShut {NoStop}%
\bibitem [{sup({\natexlab{b}})}]{supp}%
  \BibitemOpen
  \href@noop {} {}\bibinfo {howpublished}
  {\url{https://qojulia.github.io/QuantumCumulants.jl/stable/examples/unique_squeezing/}}
  ({\natexlab{b}})\BibitemShut {NoStop}%
\bibitem [{\citenamefont {Beaudoin}\ \emph {et~al.}(2011)\citenamefont
  {Beaudoin}, \citenamefont {Gambetta},\ and\ \citenamefont
  {Blais}}]{dissipation2011blais}%
  \BibitemOpen
  \bibfield  {author} {\bibinfo {author} {\bibfnamefont {F.}~\bibnamefont
  {Beaudoin}}, \bibinfo {author} {\bibfnamefont {J.~M.}\ \bibnamefont
  {Gambetta}},\ and\ \bibinfo {author} {\bibfnamefont {A.}~\bibnamefont
  {Blais}},\ }\bibfield  {title} {\bibinfo {title} {Dissipation and ultrastrong
  coupling in circuit qed},\ }\href
  {https://doi.org/10.1103/PhysRevA.84.043832} {\bibfield  {journal} {\bibinfo
  {journal} {Phys. Rev. A}\ }\textbf {\bibinfo {volume} {84}},\ \bibinfo
  {pages} {043832} (\bibinfo {year} {2011})}\BibitemShut {NoStop}%
\bibitem [{\citenamefont {Hwang}\ \emph {et~al.}(2018)\citenamefont {Hwang},
  \citenamefont {Rabl},\ and\ \citenamefont {Plenio}}]{rabl2018dissipativeQRM}%
  \BibitemOpen
  \bibfield  {author} {\bibinfo {author} {\bibfnamefont {M.-J.}\ \bibnamefont
  {Hwang}}, \bibinfo {author} {\bibfnamefont {P.}~\bibnamefont {Rabl}},\ and\
  \bibinfo {author} {\bibfnamefont {M.~B.}\ \bibnamefont {Plenio}},\ }\bibfield
   {title} {\bibinfo {title} {Dissipative phase transition in the open quantum
  rabi model},\ }\href {https://doi.org/10.1103/PhysRevA.97.013825} {\bibfield
  {journal} {\bibinfo  {journal} {Phys. Rev. A}\ }\textbf {\bibinfo {volume}
  {97}},\ \bibinfo {pages} {013825} (\bibinfo {year} {2018})}\BibitemShut
  {NoStop}%
\bibitem [{\citenamefont {Cattaneo}\ \emph {et~al.}(2019)\citenamefont
  {Cattaneo}, \citenamefont {Giorgi}, \citenamefont {Maniscalco},\ and\
  \citenamefont {Zambrini}}]{Cattaneo_2019}%
  \BibitemOpen
  \bibfield  {author} {\bibinfo {author} {\bibfnamefont {M.}~\bibnamefont
  {Cattaneo}}, \bibinfo {author} {\bibfnamefont {G.~L.}\ \bibnamefont
  {Giorgi}}, \bibinfo {author} {\bibfnamefont {S.}~\bibnamefont {Maniscalco}},\
  and\ \bibinfo {author} {\bibfnamefont {R.}~\bibnamefont {Zambrini}},\
  }\bibfield  {title} {\bibinfo {title} {Local versus global master equation
  with common and separate baths: superiority of the global approach in partial
  secular approximation},\ }\href {https://doi.org/10.1088/1367-2630/ab54ac}
  {\bibfield  {journal} {\bibinfo  {journal} {New Journal of Physics}\ }\textbf
  {\bibinfo {volume} {21}},\ \bibinfo {pages} {113045} (\bibinfo {year}
  {2019})}\BibitemShut {NoStop}%
\bibitem [{\citenamefont {Forn-D\'{\i}az}\ \emph {et~al.}(2019)\citenamefont
  {Forn-D\'{\i}az}, \citenamefont {Lamata}, \citenamefont {Rico}, \citenamefont
  {Kono},\ and\ \citenamefont {Solano}}]{RevModPhys.91.025005}%
  \BibitemOpen
  \bibfield  {author} {\bibinfo {author} {\bibfnamefont {P.}~\bibnamefont
  {Forn-D\'{\i}az}}, \bibinfo {author} {\bibfnamefont {L.}~\bibnamefont
  {Lamata}}, \bibinfo {author} {\bibfnamefont {E.}~\bibnamefont {Rico}},
  \bibinfo {author} {\bibfnamefont {J.}~\bibnamefont {Kono}},\ and\ \bibinfo
  {author} {\bibfnamefont {E.}~\bibnamefont {Solano}},\ }\bibfield  {title}
  {\bibinfo {title} {Ultrastrong coupling regimes of light-matter
  interaction},\ }\href {https://doi.org/10.1103/RevModPhys.91.025005}
  {\bibfield  {journal} {\bibinfo  {journal} {Rev. Mod. Phys.}\ }\textbf
  {\bibinfo {volume} {91}},\ \bibinfo {pages} {025005} (\bibinfo {year}
  {2019})}\BibitemShut {NoStop}%
\bibitem [{\citenamefont {Rakhubovsky}\ \emph {et~al.}(2020)\citenamefont
  {Rakhubovsky}, \citenamefont {Moore}, \citenamefont
  {Deli\ifmmode~\acute{c}\else \'{c}\fi{}}, \citenamefont {Kiesel},
  \citenamefont {Aspelmeyer},\ and\ \citenamefont {Filip}}]{uros2020}%
  \BibitemOpen
  \bibfield  {author} {\bibinfo {author} {\bibfnamefont {A.~A.}\ \bibnamefont
  {Rakhubovsky}}, \bibinfo {author} {\bibfnamefont {D.~W.}\ \bibnamefont
  {Moore}}, \bibinfo {author} {\bibfnamefont {U.~c.~v.}\ \bibnamefont
  {Deli\ifmmode~\acute{c}\else \'{c}\fi{}}}, \bibinfo {author} {\bibfnamefont
  {N.}~\bibnamefont {Kiesel}}, \bibinfo {author} {\bibfnamefont
  {M.}~\bibnamefont {Aspelmeyer}},\ and\ \bibinfo {author} {\bibfnamefont
  {R.}~\bibnamefont {Filip}},\ }\bibfield  {title} {\bibinfo {title} {Detecting
  nonclassical correlations in levitated cavity optomechanics},\ }\href
  {https://doi.org/10.1103/PhysRevApplied.14.054052} {\bibfield  {journal}
  {\bibinfo  {journal} {Phys. Rev. Appl.}\ }\textbf {\bibinfo {volume} {14}},\
  \bibinfo {pages} {054052} (\bibinfo {year} {2020})}\BibitemShut {NoStop}%
\bibitem [{\citenamefont {Rudolph}\ \emph {et~al.}(2022)\citenamefont
  {Rudolph}, \citenamefont {Deli\ifmmode~\acute{c}\else \'{c}\fi{}},
  \citenamefont {Aspelmeyer}, \citenamefont {Hornberger},\ and\ \citenamefont
  {Stickler}}]{uros2022}%
  \BibitemOpen
  \bibfield  {author} {\bibinfo {author} {\bibfnamefont {H.}~\bibnamefont
  {Rudolph}}, \bibinfo {author} {\bibfnamefont {U.~c.~v.}\ \bibnamefont
  {Deli\ifmmode~\acute{c}\else \'{c}\fi{}}}, \bibinfo {author} {\bibfnamefont
  {M.}~\bibnamefont {Aspelmeyer}}, \bibinfo {author} {\bibfnamefont
  {K.}~\bibnamefont {Hornberger}},\ and\ \bibinfo {author} {\bibfnamefont
  {B.~A.}\ \bibnamefont {Stickler}},\ }\bibfield  {title} {\bibinfo {title}
  {Force-gradient sensing and entanglement via feedback cooling of interacting
  nanoparticles},\ }\href {https://doi.org/10.1103/PhysRevLett.129.193602}
  {\bibfield  {journal} {\bibinfo  {journal} {Phys. Rev. Lett.}\ }\textbf
  {\bibinfo {volume} {129}},\ \bibinfo {pages} {193602} (\bibinfo {year}
  {2022})}\BibitemShut {NoStop}%
\bibitem [{\citenamefont {Thompson}(2015)}]{2015ionCoulombCrystal}%
  \BibitemOpen
  \bibfield  {author} {\bibinfo {author} {\bibfnamefont {R.~C.}\ \bibnamefont
  {Thompson}},\ }\bibfield  {title} {\bibinfo {title} {Ion coulomb crystals},\
  }\href {https://doi.org/10.1080/00107514.2014.989715} {\bibfield  {journal}
  {\bibinfo  {journal} {Contemp. Phys.}\ }\textbf {\bibinfo {volume} {56}},\
  \bibinfo {pages} {63} (\bibinfo {year} {2015})}\BibitemShut {NoStop}%
\bibitem [{\citenamefont {Safavi-Naini}\ \emph {et~al.}(2018)\citenamefont
  {Safavi-Naini}, \citenamefont {Lewis-Swan}, \citenamefont {Bohnet},
  \citenamefont {G\"arttner}, \citenamefont {Gilmore}, \citenamefont {Jordan},
  \citenamefont {Cohn}, \citenamefont {Freericks}, \citenamefont {Rey},\ and\
  \citenamefont {Bollinger}}]{2019CrystalDickesym}%
  \BibitemOpen
  \bibfield  {author} {\bibinfo {author} {\bibfnamefont {A.}~\bibnamefont
  {Safavi-Naini}}, \bibinfo {author} {\bibfnamefont {R.~J.}\ \bibnamefont
  {Lewis-Swan}}, \bibinfo {author} {\bibfnamefont {J.~G.}\ \bibnamefont
  {Bohnet}}, \bibinfo {author} {\bibfnamefont {M.}~\bibnamefont {G\"arttner}},
  \bibinfo {author} {\bibfnamefont {K.~A.}\ \bibnamefont {Gilmore}}, \bibinfo
  {author} {\bibfnamefont {J.~E.}\ \bibnamefont {Jordan}}, \bibinfo {author}
  {\bibfnamefont {J.}~\bibnamefont {Cohn}}, \bibinfo {author} {\bibfnamefont
  {J.~K.}\ \bibnamefont {Freericks}}, \bibinfo {author} {\bibfnamefont {A.~M.}\
  \bibnamefont {Rey}},\ and\ \bibinfo {author} {\bibfnamefont {J.~J.}\
  \bibnamefont {Bollinger}},\ }\bibfield  {title} {\bibinfo {title}
  {Verification of a many-ion simulator of the dicke model through slow
  quenches across a phase transition},\ }\href
  {https://doi.org/10.1103/PhysRevLett.121.040503} {\bibfield  {journal}
  {\bibinfo  {journal} {Phys. Rev. Lett.}\ }\textbf {\bibinfo {volume} {121}},\
  \bibinfo {pages} {040503} (\bibinfo {year} {2018})}\BibitemShut {NoStop}%
\bibitem [{\citenamefont {Hertzberg}\ \emph {et~al.}(2010)\citenamefont
  {Hertzberg}, \citenamefont {Rocheleau}, \citenamefont {Ndukum}, \citenamefont
  {Savva}, \citenamefont {Clerk},\ and\ \citenamefont
  {Schwab}}]{schwab2010nanoQBFmeasure}%
  \BibitemOpen
  \bibfield  {author} {\bibinfo {author} {\bibfnamefont {J.~B.}\ \bibnamefont
  {Hertzberg}}, \bibinfo {author} {\bibfnamefont {T.}~\bibnamefont
  {Rocheleau}}, \bibinfo {author} {\bibfnamefont {T.}~\bibnamefont {Ndukum}},
  \bibinfo {author} {\bibfnamefont {M.}~\bibnamefont {Savva}}, \bibinfo
  {author} {\bibfnamefont {A.~A.}\ \bibnamefont {Clerk}},\ and\ \bibinfo
  {author} {\bibfnamefont {K.~C.}\ \bibnamefont {Schwab}},\ }\bibfield  {title}
  {\bibinfo {title} {Back-action-evading measurements of nanomechanical
  motion},\ }\href {https://doi.org/10.1038/nphys1479} {\bibfield  {journal}
  {\bibinfo  {journal} {Nature Physics}\ }\textbf {\bibinfo {volume} {6}},\
  \bibinfo {pages} {213} (\bibinfo {year} {2010})}\BibitemShut {NoStop}%
\bibitem [{\citenamefont {M{\o}ller}\ \emph {et~al.}(2017)\citenamefont
  {M{\o}ller}, \citenamefont {Thomas}, \citenamefont {Vasilakis}, \citenamefont
  {Zeuthen}, \citenamefont {Tsaturyan}, \citenamefont {Balabas}, \citenamefont
  {Jensen}, \citenamefont {Schliesser}, \citenamefont {Hammerer},\ and\
  \citenamefont {Polzik}}]{polzik2017baemeasure}%
  \BibitemOpen
  \bibfield  {author} {\bibinfo {author} {\bibfnamefont {C.~B.}\ \bibnamefont
  {M{\o}ller}}, \bibinfo {author} {\bibfnamefont {R.~A.}\ \bibnamefont
  {Thomas}}, \bibinfo {author} {\bibfnamefont {G.}~\bibnamefont {Vasilakis}},
  \bibinfo {author} {\bibfnamefont {E.}~\bibnamefont {Zeuthen}}, \bibinfo
  {author} {\bibfnamefont {Y.}~\bibnamefont {Tsaturyan}}, \bibinfo {author}
  {\bibfnamefont {M.}~\bibnamefont {Balabas}}, \bibinfo {author} {\bibfnamefont
  {K.}~\bibnamefont {Jensen}}, \bibinfo {author} {\bibfnamefont
  {A.}~\bibnamefont {Schliesser}}, \bibinfo {author} {\bibfnamefont
  {K.}~\bibnamefont {Hammerer}},\ and\ \bibinfo {author} {\bibfnamefont
  {E.~S.}\ \bibnamefont {Polzik}},\ }\bibfield  {title} {\bibinfo {title}
  {Quantum back-action-evading measurement of motion in a negative mass
  reference frame},\ }\href {https://doi.org/10.1038/nature22980} {\bibfield
  {journal} {\bibinfo  {journal} {Nature}\ }\textbf {\bibinfo {volume} {547}},\
  \bibinfo {pages} {191} (\bibinfo {year} {2017})}\BibitemShut {NoStop}%
\bibitem [{\citenamefont {Pavlov}\ \emph {et~al.}(2023)\citenamefont {Pavlov},
  \citenamefont {Porras},\ and\ \citenamefont
  {Ivanov}}]{porras2023ddmetrologycollectivespin}%
  \BibitemOpen
  \bibfield  {author} {\bibinfo {author} {\bibfnamefont {V.~P.}\ \bibnamefont
  {Pavlov}}, \bibinfo {author} {\bibfnamefont {D.}~\bibnamefont {Porras}},\
  and\ \bibinfo {author} {\bibfnamefont {P.~A.}\ \bibnamefont {Ivanov}},\
  }\href@noop {} {\bibinfo {title} {Quantum metrology with critical
  driven-dissipative collective spin system}} (\bibinfo {year} {2023}),\
  \Eprint {https://arxiv.org/abs/2302.05216} {arXiv:2302.05216 [quant-ph]}
  \BibitemShut {NoStop}%
\bibitem [{\citenamefont {Ilias}\ \emph {et~al.}(2023)\citenamefont {Ilias},
  \citenamefont {Yang}, \citenamefont {Huelga},\ and\ \citenamefont
  {Plenio}}]{ilias2023criticalityenhanced}%
  \BibitemOpen
  \bibfield  {author} {\bibinfo {author} {\bibfnamefont {T.}~\bibnamefont
  {Ilias}}, \bibinfo {author} {\bibfnamefont {D.}~\bibnamefont {Yang}},
  \bibinfo {author} {\bibfnamefont {S.~F.}\ \bibnamefont {Huelga}},\ and\
  \bibinfo {author} {\bibfnamefont {M.~B.}\ \bibnamefont {Plenio}},\
  }\href@noop {} {\bibinfo {title} {Criticality-enhanced electromagnetic field
  sensor with single trapped ions}} (\bibinfo {year} {2023}),\ \Eprint
  {https://arxiv.org/abs/2304.02050} {arXiv:2304.02050 [quant-ph]} \BibitemShut
  {NoStop}%
\bibitem [{\citenamefont {Mirkhalaf}\ \emph {et~al.}(2020)\citenamefont
  {Mirkhalaf}, \citenamefont {Witkowska},\ and\ \citenamefont
  {Lepori}}]{2020criticalwitkowska}%
  \BibitemOpen
  \bibfield  {author} {\bibinfo {author} {\bibfnamefont {S.~S.}\ \bibnamefont
  {Mirkhalaf}}, \bibinfo {author} {\bibfnamefont {E.}~\bibnamefont
  {Witkowska}},\ and\ \bibinfo {author} {\bibfnamefont {L.}~\bibnamefont
  {Lepori}},\ }\bibfield  {title} {\bibinfo {title} {Supersensitive quantum
  sensor based on criticality in an antiferromagnetic spinor condensate},\
  }\href {https://doi.org/10.1103/PhysRevA.101.043609} {\bibfield  {journal}
  {\bibinfo  {journal} {Phys. Rev. A}\ }\textbf {\bibinfo {volume} {101}},\
  \bibinfo {pages} {043609} (\bibinfo {year} {2020})}\BibitemShut {NoStop}%
\bibitem [{\citenamefont {Gietka}\ \emph {et~al.}(2021)\citenamefont {Gietka},
  \citenamefont {Metz}, \citenamefont {Keller},\ and\ \citenamefont
  {Li}}]{Gietka2021adiabaticcritical}%
  \BibitemOpen
  \bibfield  {author} {\bibinfo {author} {\bibfnamefont {K.}~\bibnamefont
  {Gietka}}, \bibinfo {author} {\bibfnamefont {F.}~\bibnamefont {Metz}},
  \bibinfo {author} {\bibfnamefont {T.}~\bibnamefont {Keller}},\ and\ \bibinfo
  {author} {\bibfnamefont {J.}~\bibnamefont {Li}},\ }\bibfield  {title}
  {\bibinfo {title} {Adiabatic critical quantum metrology cannot reach the
  {H}eisenberg limit even when shortcuts to adiabaticity are applied},\ }\href
  {https://doi.org/10.22331/q-2021-07-01-489} {\bibfield  {journal} {\bibinfo
  {journal} {{Quantum}}\ }\textbf {\bibinfo {volume} {5}},\ \bibinfo {pages}
  {489} (\bibinfo {year} {2021})}\BibitemShut {NoStop}%
\bibitem [{\citenamefont {Liu}\ \emph {et~al.}(2021)\citenamefont {Liu},
  \citenamefont {Chen}, \citenamefont {Jiang}, \citenamefont {Yang},
  \citenamefont {Wu}, \citenamefont {Li}, \citenamefont {Yuan}, \citenamefont
  {Peng},\ and\ \citenamefont {Du}}]{2021liuexperiment}%
  \BibitemOpen
  \bibfield  {author} {\bibinfo {author} {\bibfnamefont {R.}~\bibnamefont
  {Liu}}, \bibinfo {author} {\bibfnamefont {Y.}~\bibnamefont {Chen}}, \bibinfo
  {author} {\bibfnamefont {M.}~\bibnamefont {Jiang}}, \bibinfo {author}
  {\bibfnamefont {X.}~\bibnamefont {Yang}}, \bibinfo {author} {\bibfnamefont
  {Z.}~\bibnamefont {Wu}}, \bibinfo {author} {\bibfnamefont {Y.}~\bibnamefont
  {Li}}, \bibinfo {author} {\bibfnamefont {H.}~\bibnamefont {Yuan}}, \bibinfo
  {author} {\bibfnamefont {X.}~\bibnamefont {Peng}},\ and\ \bibinfo {author}
  {\bibfnamefont {J.}~\bibnamefont {Du}},\ }\bibfield  {title} {\bibinfo
  {title} {Experimental critical quantum metrology with the heisenberg
  scaling},\ }\href {https://doi.org/10.1038/s41534-021-00507-x} {\bibfield
  {journal} {\bibinfo  {journal} {npj Quantum Information}\ }\textbf {\bibinfo
  {volume} {7}},\ \bibinfo {pages} {170} (\bibinfo {year} {2021})}\BibitemShut
  {NoStop}%
\bibitem [{\citenamefont {Mirkhalaf}\ \emph {et~al.}(2021)\citenamefont
  {Mirkhalaf}, \citenamefont {Benedicto~Orenes}, \citenamefont {Mitchell},\
  and\ \citenamefont {Witkowska}}]{2021criticalwitkowska}%
  \BibitemOpen
  \bibfield  {author} {\bibinfo {author} {\bibfnamefont {S.~S.}\ \bibnamefont
  {Mirkhalaf}}, \bibinfo {author} {\bibfnamefont {D.}~\bibnamefont
  {Benedicto~Orenes}}, \bibinfo {author} {\bibfnamefont {M.~W.}\ \bibnamefont
  {Mitchell}},\ and\ \bibinfo {author} {\bibfnamefont {E.}~\bibnamefont
  {Witkowska}},\ }\bibfield  {title} {\bibinfo {title} {Criticality-enhanced
  quantum sensing in ferromagnetic bose-einstein condensates: Role of readout
  measurement and detection noise},\ }\href
  {https://doi.org/10.1103/PhysRevA.103.023317} {\bibfield  {journal} {\bibinfo
   {journal} {Phys. Rev. A}\ }\textbf {\bibinfo {volume} {103}},\ \bibinfo
  {pages} {023317} (\bibinfo {year} {2021})}\BibitemShut {NoStop}%
\bibitem [{\citenamefont {Ilias}\ \emph {et~al.}(2022)\citenamefont {Ilias},
  \citenamefont {Yang}, \citenamefont {Huelga},\ and\ \citenamefont
  {Plenio}}]{2022PRXQuantumcontinus}%
  \BibitemOpen
  \bibfield  {author} {\bibinfo {author} {\bibfnamefont {T.}~\bibnamefont
  {Ilias}}, \bibinfo {author} {\bibfnamefont {D.}~\bibnamefont {Yang}},
  \bibinfo {author} {\bibfnamefont {S.~F.}\ \bibnamefont {Huelga}},\ and\
  \bibinfo {author} {\bibfnamefont {M.~B.}\ \bibnamefont {Plenio}},\ }\bibfield
   {title} {\bibinfo {title} {Criticality-enhanced quantum sensing via
  continuous measurement},\ }\href
  {https://doi.org/10.1103/PRXQuantum.3.010354} {\bibfield  {journal} {\bibinfo
   {journal} {PRX Quantum}\ }\textbf {\bibinfo {volume} {3}},\ \bibinfo {pages}
  {010354} (\bibinfo {year} {2022})}\BibitemShut {NoStop}%
\bibitem [{\citenamefont {Garbe}\ \emph
  {et~al.}(2022{\natexlab{a}})\citenamefont {Garbe}, \citenamefont {Abah},
  \citenamefont {Felicetti},\ and\ \citenamefont
  {Puebla}}]{2022_Garbe_heisenbegkible}%
  \BibitemOpen
  \bibfield  {author} {\bibinfo {author} {\bibfnamefont {L.}~\bibnamefont
  {Garbe}}, \bibinfo {author} {\bibfnamefont {O.}~\bibnamefont {Abah}},
  \bibinfo {author} {\bibfnamefont {S.}~\bibnamefont {Felicetti}},\ and\
  \bibinfo {author} {\bibfnamefont {R.}~\bibnamefont {Puebla}},\ }\bibfield
  {title} {\bibinfo {title} {Critical quantum metrology with fully-connected
  models: from heisenberg to kibble–zurek scaling},\ }\href
  {https://doi.org/10.1088/2058-9565/ac6ca5} {\bibfield  {journal} {\bibinfo
  {journal} {Quantum Science and Technology}\ }\textbf {\bibinfo {volume}
  {7}},\ \bibinfo {pages} {035010} (\bibinfo {year}
  {2022}{\natexlab{a}})}\BibitemShut {NoStop}%
\bibitem [{\citenamefont {Gietka}\ \emph {et~al.}(2022)\citenamefont {Gietka},
  \citenamefont {Ruks},\ and\ \citenamefont {Busch}}]{Gietka2022understanding}%
  \BibitemOpen
  \bibfield  {author} {\bibinfo {author} {\bibfnamefont {K.}~\bibnamefont
  {Gietka}}, \bibinfo {author} {\bibfnamefont {L.}~\bibnamefont {Ruks}},\ and\
  \bibinfo {author} {\bibfnamefont {T.}~\bibnamefont {Busch}},\ }\bibfield
  {title} {\bibinfo {title} {Understanding and {I}mproving {C}ritical
  {M}etrology. {Q}uenching {S}uperradiant {L}ight-{M}atter {S}ystems {B}eyond
  the {C}ritical {P}oint},\ }\href {https://doi.org/10.22331/q-2022-04-27-700}
  {\bibfield  {journal} {\bibinfo  {journal} {{Quantum}}\ }\textbf {\bibinfo
  {volume} {6}},\ \bibinfo {pages} {700} (\bibinfo {year} {2022})}\BibitemShut
  {NoStop}%
\bibitem [{\citenamefont {Ying}\ \emph {et~al.}(2022)\citenamefont {Ying},
  \citenamefont {Felicetti}, \citenamefont {Liu},\ and\ \citenamefont
  {Braak}}]{2022entropycritical}%
  \BibitemOpen
  \bibfield  {author} {\bibinfo {author} {\bibfnamefont {Z.-J.}\ \bibnamefont
  {Ying}}, \bibinfo {author} {\bibfnamefont {S.}~\bibnamefont {Felicetti}},
  \bibinfo {author} {\bibfnamefont {G.}~\bibnamefont {Liu}},\ and\ \bibinfo
  {author} {\bibfnamefont {D.}~\bibnamefont {Braak}},\ }\bibfield  {title}
  {\bibinfo {title} {Critical quantum metrology in the non-linear quantum rabi
  model},\ }\href {https://doi.org/10.3390/e24081015} {\bibfield  {journal}
  {\bibinfo  {journal} {Entropy}\ }\textbf {\bibinfo {volume} {24}} (\bibinfo
  {year} {2022})}\BibitemShut {NoStop}%
\bibitem [{\citenamefont {Ding}\ \emph {et~al.}(2022)\citenamefont {Ding},
  \citenamefont {Liu}, \citenamefont {Shi}, \citenamefont {Guo}, \citenamefont
  {M{\o}lmer},\ and\ \citenamefont {Adams}}]{2022dingnature}%
  \BibitemOpen
  \bibfield  {author} {\bibinfo {author} {\bibfnamefont {D.-S.}\ \bibnamefont
  {Ding}}, \bibinfo {author} {\bibfnamefont {Z.-K.}\ \bibnamefont {Liu}},
  \bibinfo {author} {\bibfnamefont {B.-S.}\ \bibnamefont {Shi}}, \bibinfo
  {author} {\bibfnamefont {G.-C.}\ \bibnamefont {Guo}}, \bibinfo {author}
  {\bibfnamefont {K.}~\bibnamefont {M{\o}lmer}},\ and\ \bibinfo {author}
  {\bibfnamefont {C.~S.}\ \bibnamefont {Adams}},\ }\bibfield  {title} {\bibinfo
  {title} {Enhanced metrology at the critical point of a many-body rydberg
  atomic system},\ }\href {https://doi.org/10.1038/s41567-022-01777-8}
  {\bibfield  {journal} {\bibinfo  {journal} {Nature Physics}\ } (\bibinfo
  {year} {2022})}\BibitemShut {NoStop}%
\bibitem [{\citenamefont {Aybar}\ \emph {et~al.}(2022)\citenamefont {Aybar},
  \citenamefont {Niezgoda}, \citenamefont {Mirkhalaf}, \citenamefont
  {Mitchell}, \citenamefont {Benedicto~Orenes},\ and\ \citenamefont
  {Witkowska}}]{Aybar2022criticalquantum}%
  \BibitemOpen
  \bibfield  {author} {\bibinfo {author} {\bibfnamefont {E.}~\bibnamefont
  {Aybar}}, \bibinfo {author} {\bibfnamefont {A.}~\bibnamefont {Niezgoda}},
  \bibinfo {author} {\bibfnamefont {S.~S.}\ \bibnamefont {Mirkhalaf}}, \bibinfo
  {author} {\bibfnamefont {M.~W.}\ \bibnamefont {Mitchell}}, \bibinfo {author}
  {\bibfnamefont {D.}~\bibnamefont {Benedicto~Orenes}},\ and\ \bibinfo {author}
  {\bibfnamefont {E.}~\bibnamefont {Witkowska}},\ }\bibfield  {title} {\bibinfo
  {title} {Critical quantum thermometry and its feasibility in spin systems},\
  }\href {https://doi.org/10.22331/q-2022-09-19-808} {\bibfield  {journal}
  {\bibinfo  {journal} {{Quantum}}\ }\textbf {\bibinfo {volume} {6}},\ \bibinfo
  {pages} {808} (\bibinfo {year} {2022})}\BibitemShut {NoStop}%
\bibitem [{\citenamefont {Garbe}\ \emph
  {et~al.}(2022{\natexlab{b}})\citenamefont {Garbe}, \citenamefont {Abah},
  \citenamefont {Felicetti},\ and\ \citenamefont
  {Puebla}}]{2022garbeexponetialsqueezing}%
  \BibitemOpen
  \bibfield  {author} {\bibinfo {author} {\bibfnamefont {L.}~\bibnamefont
  {Garbe}}, \bibinfo {author} {\bibfnamefont {O.}~\bibnamefont {Abah}},
  \bibinfo {author} {\bibfnamefont {S.}~\bibnamefont {Felicetti}},\ and\
  \bibinfo {author} {\bibfnamefont {R.}~\bibnamefont {Puebla}},\ }\bibfield
  {title} {\bibinfo {title} {Exponential time-scaling of estimation precision
  by reaching a quantum critical point},\ }\href
  {https://doi.org/10.1103/PhysRevResearch.4.043061} {\bibfield  {journal}
  {\bibinfo  {journal} {Phys. Rev. Research}\ }\textbf {\bibinfo {volume}
  {4}},\ \bibinfo {pages} {043061} (\bibinfo {year}
  {2022}{\natexlab{b}})}\BibitemShut {NoStop}%
\bibitem [{\citenamefont {Gietka}\ and\ \citenamefont
  {Ritsch}(2023)}]{gietka2023squeezingHeisenberg}%
  \BibitemOpen
  \bibfield  {author} {\bibinfo {author} {\bibfnamefont {K.}~\bibnamefont
  {Gietka}}\ and\ \bibinfo {author} {\bibfnamefont {H.}~\bibnamefont
  {Ritsch}},\ }\bibfield  {title} {\bibinfo {title} {Squeezing and overcoming
  the heisenberg scaling with spin-orbit coupled quantum gases},\ }\href
  {https://doi.org/10.1103/PhysRevLett.130.090802} {\bibfield  {journal}
  {\bibinfo  {journal} {Phys. Rev. Lett.}\ }\textbf {\bibinfo {volume} {130}},\
  \bibinfo {pages} {090802} (\bibinfo {year} {2023})}\BibitemShut {NoStop}%
\bibitem [{\citenamefont {Kronwald}\ \emph {et~al.}(2013)\citenamefont
  {Kronwald}, \citenamefont {Marquardt},\ and\ \citenamefont
  {Clerk}}]{2013clerk}%
  \BibitemOpen
  \bibfield  {author} {\bibinfo {author} {\bibfnamefont {A.}~\bibnamefont
  {Kronwald}}, \bibinfo {author} {\bibfnamefont {F.}~\bibnamefont
  {Marquardt}},\ and\ \bibinfo {author} {\bibfnamefont {A.~A.}\ \bibnamefont
  {Clerk}},\ }\bibfield  {title} {\bibinfo {title} {Arbitrarily large
  steady-state bosonic squeezing via dissipation},\ }\href
  {https://doi.org/10.1103/PhysRevA.88.063833} {\bibfield  {journal} {\bibinfo
  {journal} {Phys. Rev. A}\ }\textbf {\bibinfo {volume} {88}},\ \bibinfo
  {pages} {063833} (\bibinfo {year} {2013})}\BibitemShut {NoStop}%
\bibitem [{\citenamefont {Aspelmeyer}\ \emph {et~al.}(2014)\citenamefont
  {Aspelmeyer}, \citenamefont {Kippenberg},\ and\ \citenamefont
  {Marquardt}}]{aspelmeyer2014cavityoptom}%
  \BibitemOpen
  \bibfield  {author} {\bibinfo {author} {\bibfnamefont {M.}~\bibnamefont
  {Aspelmeyer}}, \bibinfo {author} {\bibfnamefont {T.~J.}\ \bibnamefont
  {Kippenberg}},\ and\ \bibinfo {author} {\bibfnamefont {F.}~\bibnamefont
  {Marquardt}},\ }\bibfield  {title} {\bibinfo {title} {Cavity optomechanics},\
  }\href {https://doi.org/10.1103/RevModPhys.86.1391} {\bibfield  {journal}
  {\bibinfo  {journal} {Rev. Mod. Phys.}\ }\textbf {\bibinfo {volume} {86}},\
  \bibinfo {pages} {1391} (\bibinfo {year} {2014})}\BibitemShut {NoStop}%
\bibitem [{\citenamefont {Dare}\ \emph {et~al.}(2023)\citenamefont {Dare},
  \citenamefont {Hansen}, \citenamefont {Coroli}, \citenamefont {Johnson},
  \citenamefont {Aspelmeyer},\ and\ \citenamefont {Delić}}]{dare2023linear}%
  \BibitemOpen
  \bibfield  {author} {\bibinfo {author} {\bibfnamefont {K.}~\bibnamefont
  {Dare}}, \bibinfo {author} {\bibfnamefont {J.~J.}\ \bibnamefont {Hansen}},
  \bibinfo {author} {\bibfnamefont {I.}~\bibnamefont {Coroli}}, \bibinfo
  {author} {\bibfnamefont {A.}~\bibnamefont {Johnson}}, \bibinfo {author}
  {\bibfnamefont {M.}~\bibnamefont {Aspelmeyer}},\ and\ \bibinfo {author}
  {\bibfnamefont {U.}~\bibnamefont {Delić}},\ }\href@noop {} {\bibinfo {title}
  {Linear ultrastrong optomechanical interaction}} (\bibinfo {year} {2023}),\
  \Eprint {https://arxiv.org/abs/2305.16226} {arXiv:2305.16226 [quant-ph]}
  \BibitemShut {NoStop}%
\bibitem [{\citenamefont {Zhang}\ \emph {et~al.}(2016)\citenamefont {Zhang},
  \citenamefont {Mossman}, \citenamefont {Busch}, \citenamefont {Engels},\ and\
  \citenamefont {Zhang}}]{busch2016socreview}%
  \BibitemOpen
  \bibfield  {author} {\bibinfo {author} {\bibfnamefont {Y.}~\bibnamefont
  {Zhang}}, \bibinfo {author} {\bibfnamefont {M.~E.}\ \bibnamefont {Mossman}},
  \bibinfo {author} {\bibfnamefont {T.}~\bibnamefont {Busch}}, \bibinfo
  {author} {\bibfnamefont {P.}~\bibnamefont {Engels}},\ and\ \bibinfo {author}
  {\bibfnamefont {C.}~\bibnamefont {Zhang}},\ }\bibfield  {title} {\bibinfo
  {title} {Properties of spin--orbit-coupled bose--einstein condensates},\
  }\href {https://doi.org/10.1007/s11467-016-0560-y} {\bibfield  {journal}
  {\bibinfo  {journal} {Frontiers of Physics}\ }\textbf {\bibinfo {volume}
  {11}},\ \bibinfo {pages} {118103} (\bibinfo {year} {2016})}\BibitemShut
  {NoStop}%
\bibitem [{\citenamefont {Cai}\ \emph {et~al.}(2021)\citenamefont {Cai},
  \citenamefont {Liu}, \citenamefont {Zhao}, \citenamefont {Wu}, \citenamefont
  {Mei}, \citenamefont {Jiang}, \citenamefont {He}, \citenamefont {Zhang},
  \citenamefont {Zhou},\ and\ \citenamefont {Duan}}]{duan2021qrbphonon}%
  \BibitemOpen
  \bibfield  {author} {\bibinfo {author} {\bibfnamefont {M.~L.}\ \bibnamefont
  {Cai}}, \bibinfo {author} {\bibfnamefont {Z.~D.}\ \bibnamefont {Liu}},
  \bibinfo {author} {\bibfnamefont {W.~D.}\ \bibnamefont {Zhao}}, \bibinfo
  {author} {\bibfnamefont {Y.~K.}\ \bibnamefont {Wu}}, \bibinfo {author}
  {\bibfnamefont {Q.~X.}\ \bibnamefont {Mei}}, \bibinfo {author} {\bibfnamefont
  {Y.}~\bibnamefont {Jiang}}, \bibinfo {author} {\bibfnamefont
  {L.}~\bibnamefont {He}}, \bibinfo {author} {\bibfnamefont {X.}~\bibnamefont
  {Zhang}}, \bibinfo {author} {\bibfnamefont {Z.~C.}\ \bibnamefont {Zhou}},\
  and\ \bibinfo {author} {\bibfnamefont {L.~M.}\ \bibnamefont {Duan}},\
  }\bibfield  {title} {\bibinfo {title} {Observation of a quantum phase
  transition in the quantum rabi model with a single trapped ion},\ }\href
  {https://doi.org/10.1038/s41467-021-21425-8} {\bibfield  {journal} {\bibinfo
  {journal} {Nature Communications}\ }\textbf {\bibinfo {volume} {12}},\
  \bibinfo {pages} {1126} (\bibinfo {year} {2021})}\BibitemShut {NoStop}%
\bibitem [{\citenamefont {Yunusova}\ \emph {et~al.}(2019)\citenamefont
  {Yunusova}, \citenamefont {Konstantinov}, \citenamefont {Bouchiat},\ and\
  \citenamefont {Chepelianskii}}]{konstantinov2019qrmhelium}%
  \BibitemOpen
  \bibfield  {author} {\bibinfo {author} {\bibfnamefont {K.~M.}\ \bibnamefont
  {Yunusova}}, \bibinfo {author} {\bibfnamefont {D.}~\bibnamefont
  {Konstantinov}}, \bibinfo {author} {\bibfnamefont {H.}~\bibnamefont
  {Bouchiat}},\ and\ \bibinfo {author} {\bibfnamefont {A.~D.}\ \bibnamefont
  {Chepelianskii}},\ }\bibfield  {title} {\bibinfo {title} {Coupling between
  rydberg states and landau levels of electrons trapped on liquid helium},\
  }\href {https://doi.org/10.1103/PhysRevLett.122.176802} {\bibfield  {journal}
  {\bibinfo  {journal} {Phys. Rev. Lett.}\ }\textbf {\bibinfo {volume} {122}},\
  \bibinfo {pages} {176802} (\bibinfo {year} {2019})}\BibitemShut {NoStop}%
\bibitem [{\citenamefont {Kustura}\ \emph {et~al.}(2022)\citenamefont
  {Kustura}, \citenamefont {Gonzalez-Ballestero}, \citenamefont {Sommer},
  \citenamefont {Meyer}, \citenamefont {Quidant},\ and\ \citenamefont
  {Romero-Isart}}]{kustura2022mechanicalsqueezing}%
  \BibitemOpen
  \bibfield  {author} {\bibinfo {author} {\bibfnamefont {K.}~\bibnamefont
  {Kustura}}, \bibinfo {author} {\bibfnamefont {C.}~\bibnamefont
  {Gonzalez-Ballestero}}, \bibinfo {author} {\bibfnamefont {A.~d. l.~R.}\
  \bibnamefont {Sommer}}, \bibinfo {author} {\bibfnamefont {N.}~\bibnamefont
  {Meyer}}, \bibinfo {author} {\bibfnamefont {R.}~\bibnamefont {Quidant}},\
  and\ \bibinfo {author} {\bibfnamefont {O.}~\bibnamefont {Romero-Isart}},\
  }\bibfield  {title} {\bibinfo {title} {Mechanical squeezing via unstable
  dynamics in a microcavity},\ }\href
  {https://doi.org/10.1103/PhysRevLett.128.143601} {\bibfield  {journal}
  {\bibinfo  {journal} {Phys. Rev. Lett.}\ }\textbf {\bibinfo {volume} {128}},\
  \bibinfo {pages} {143601} (\bibinfo {year} {2022})}\BibitemShut {NoStop}%
\bibitem [{\citenamefont {Kr\"{a}mer}\ \emph {et~al.}(2018)\citenamefont
  {Kr\"{a}mer}, \citenamefont {Plankensteiner}, \citenamefont {Ostermann},\
  and\ \citenamefont {Ritsch}}]{kramer2018quantumoptics}%
  \BibitemOpen
  \bibfield  {author} {\bibinfo {author} {\bibfnamefont {S.}~\bibnamefont
  {Kr\"{a}mer}}, \bibinfo {author} {\bibfnamefont {D.}~\bibnamefont
  {Plankensteiner}}, \bibinfo {author} {\bibfnamefont {L.}~\bibnamefont
  {Ostermann}},\ and\ \bibinfo {author} {\bibfnamefont {H.}~\bibnamefont
  {Ritsch}},\ }\bibfield  {title} {\bibinfo {title} {{QuantumOptics}.jl: A
  {J}ulia framework for simulating open quantum systems},\ }\href
  {https://dx.doi.org/10.1016/j.cpc.2018.02.004} {\bibfield  {journal}
  {\bibinfo  {journal} {Comput. Phys. Commun}\ }\textbf {\bibinfo {volume}
  {227}} (\bibinfo {year} {2018})}\BibitemShut {NoStop}%
\end{thebibliography}


\end{document}